\documentclass{article}
\usepackage{graphics,amsmath,amsfonts,amscd,revsymb,latexsym, enumerate}
\newtheorem{theorem}{Theorem}[section]

\newtheorem{lemma}[theorem]{Lemma}

\newtheorem{proposition}[theorem]{Proposition}

\newcommand{\qed}{{\hfill$\Box$}}
\newenvironment{proof}{\noindent \textbf{{Proof~} }}{\qed}

\def\bi{\begin{itemize}}
\def\ei{\end{itemize}}
\def\be{\begin{equation}}
\def\ee{\end{equation}}
\def\bea{\begin{eqnarray}}
\def\eea{\end{eqnarray}}
\def\ben{\begin{eqnarray*}}
\def\een{\end{eqnarray*}}

\def\>{\rangle}
\def\<{\langle}

\def\br{{\bf r}}

\def\H{F}

\newcommand{\bA}{{\mathbf A}}
\newcommand{\bB}{{\mathbf B}}

\newcommand{\bI}{{\mathbf I}}

\newcommand{\bu}{{\mathbf u}}
\newcommand{\bv}{{\mathbf v}}
\newcommand{\bx}{{\mathbf x}}
\newcommand{\bz}{{\mathbf z}}
\newcommand{\by}{{\mathbf y}}
\newcommand{\bw}{{\mathbf w}}
\newcommand{\ba}{{\mathbf a}}
\newcommand{\bb}{{\mathbf b}}
\newcommand{\bc}{{\mathbf c}}
\newcommand{\bg}{{\mathbf g}}
\newcommand{\bh}{{\mathbf h}}

\newcommand{\bo}{{\mathbf 0}}
\newcommand{\bee}{{\mathbf e}}
\newcommand{\bal}{{\mathbf \alpha}}
\newcommand{\bbe}{{\mathbf \beta}}

\def\bbC{\mathbb{C}}

\def\bbZ{\mathbb{Z}}
\def\bbF{\mathbb{F}}
\newcommand{\1} I %{{\openone}}

\def\T{{{T}}}

\newcommand{\inner}[2]{\langle{#1},{#2}\rangle}
\newcommand{\bra}[1]{\langle #1 |}
\newcommand{\ket}[1]{| #1 \rangle}

\newcommand{\proj}[1]{| #1 \>\!\< #1 |}

\newcommand{\rs}[1] {{\rm rowspace} ( #1)}
\newcommand{\iso}[1] {{\rm iso} ( #1)}
\newcommand{\symp}[1] {{\rm symp} ( #1)}
\newcommand{\spann}[1] {{\rm span} \{ #1 \}}

\DeclareMathOperator{\id}{id}

\DeclareMathOperator{\tr}{Tr}

\DeclareMathOperator{\wt}{wt}

\def\*{\star}

\def\tilde{\widetilde}
\def\bar{\overline}

\def\cC{{\cal C}}
\def\cD{{\cal D}}        \def\cE{{\cal E}}

\def\cG{{\cal G}}        \def\cH{{\cal H}}

\def\cL{{\cal L}}

        \def\cN{{\cal N}}

\def\cR{{\cal R}}
        
\def\cU{{\cal U}}
\def\cV{{\cal V}}

\begin{document}

\title{Catalytic quantum error correction
\protect\vspace{1cm}}

\author{Todd Brun\footnote{\tt tbrun@usc.edu}, 
Igor Devetak\footnote{\tt devetak@usc.edu}
  \, and Min-Hsiu Hsieh\footnote{\tt minhsiuh@usc.edu} \protect\\
\\
\it{Department of Electrical Engineering--Systems,}
\it{University of Southern California,}\protect\\
\it{Los Angeles, CA 90089, USA}\protect\\
}

%\date{\small (Dated 30 April 2005)}
\date{\today}

\maketitle

\begin{abstract}
We develop the theory of \emph{entanglement-assisted 
quantum error correcting} (EAQEC)
\emph{codes}, a generalization of the stabilizer formalism 
to the setting in which the sender and receiver have access to pre-shared 
entanglement.  
Conventional stabilizer codes are equivalent to dual-containing
symplectic codes. In contrast, EAQEC codes do not
require the dual-containing condition, which greatly simplifies their construction.
We show how any quaternary classical code
can be made into a EAQEC code. In particular, efficient modern
codes, like LDPC codes, which attain the Shannon capacity,
can be made into EAQEC codes attaining the
hashing bound. In a quantum computation setting,
EAQEC codes give rise to \emph{catalytic} quantum codes
which  maintain a region of inherited noiseless qubits.
 We also give an alternative construction of
EAQEC codes by making classical entanglement assisted codes
coherent.

\end{abstract}

%The most important feature of our construction is that
%convert any classical $[n,k,d]$ symplectic code into the equivalent of
%an $[[n,2k - n,d]]$ quantum error-correcting code.
%In contrast, the conventional stabilizer formalism
%imposes self-orthogonality onto the corresponding classical symplectic codes.
%This should be contrasted
%to the conventional stabilizer formalism which corresponds to
%\end{abstract}

Information theory and the theory of error-correcting codes
(coding theory) are intimately connected. Both
address  the problem of sending information over noisy channels.
The sender Alice encodes her message as a codeword, sends it through the channel,
and the receiver Bob tries to infer the intended message based on the channel output.

Information theory (or rather the subfield of Shannon theory)
deals with the \emph{asymptotic} setting of increasingly
long codes, with asymptotically vanishing error probability. The noisy channel
is typically assumed to act independently on the codeword bits.
The fundamental quantity of interest is the \emph{capacity} of the channel:
the optimal rate (in bits per channel use) of information transfer.
Claude Shannon  \cite{Shannon48} gave a remarkable characterization of the channel
capacity in terms of mutual information. Unfortunately, the capacity
is achieved by random coding, which means highly inefficient encoding and decoding
algorithms.

Coding theory deals with the practical \emph{finite} setting, characterized by
a fixed code length, number of encoded bits and correctable error set.
%It is concerned with codes with \emph{structure}, to
%ensure efficient encoding and decoding.
The most popular codes have simple mathematical properties,
such as linearity (a linear combination of codewords is another codeword),
which allows for efficient encoding.
The performance of these codes is then measured against the optimal performance
set by Shannon theory.

This relationship carries over to quantum information processing.
The basic communication task is sending quantum information over
noisy quantum channels. This setting is also relevant for
fault tolerant quantum computation, because decoherence can be regarded
as a quantum channel connecting two points in time (rather than space).
The first quantum error-correcting (QEC) code  was discovered by Shor
\cite{Sho95}, leading to an explosion of research in
subsequent  years \cite{Got96,BDSW96,LMPZ96,KL97,Ste96,CS96,Got98,CRSS97,
CRSS98}. Calderbank and Shor \cite{CS96} and Steane \cite{Ste96}
gave the first systematic way to construct quantum ``CSS'' codes
from dual-containing classical
codes over $\bbZ_2$. These efforts culminated in a general theory of linear quantum codes,
also known as \emph{stabilizer codes} \cite{CRSS98, Got98, Pre98, NC00}.
Stabilizer codes are equivalent to classical codes which are dual-containing with respect
to the symplectic bilinear form. These in turn may be constructed
from dual-containing classical codes over $\bbF_4$, % (also known as GF(4)),
generalizing the CSS construction \cite{CRSS97}.

In  \cite{Sho95} Shor also raised the information theoretical question of
characterizing the capacity of a quantum channel for sending quantum information,
 subsequently answered by
\cite{Lloyd96, Shor02, Devetak03} in terms of  \emph{coherent information}.
%Quantum information theory gives us asymptotic
%bounds which we are trying to achieve with quantum codes.
It comes as no surprise that coding theory and information theory
continue to inform each other in the quantum setting.
%have a lot to say to each other.
The capacity-achieving quantum codes of \cite{Devetak03} have a
structure akin to CSS codes (thanks to their common connection to cryptography).
Concatenated stabilizer codes achieve rates equal to the coherent information evaluated on
density operators corresponding to maximally mixed qubit states encoded by a stablizer code
\cite{BDSW96, Hamada02}.

Research has since taken us beyond this most obvious
quantum communication setting. Apart from quantum communication channels,
there are other resources to consider, such as entanglement and classical communication.
Great progress has been made in characterizing optimal tradeoffs between these resources.
For example, the capacity of a quantum channel for sending classical information
assisted by entanglement (EA capacity)
is a simple single letter expression involving quantum mutual information
\cite{BSST01}. In \cite{DHW03} a  remarkable duality was discovered
 between entanglement-assisted quantum communication (the ``father'' protocol)
and quantum-communication-assisted entanglement distillation (the ``mother'' protocol).
The two were shown to generate a whole family of protocols
when combined with the more elementary protocols of superdense coding \cite{BW92},
quantum teleportation \cite{BBCJPW93} and entanglement distribution \cite{DHW03}.

\begin{figure}
\centerline{ {\scalebox{0.60}{\includegraphics{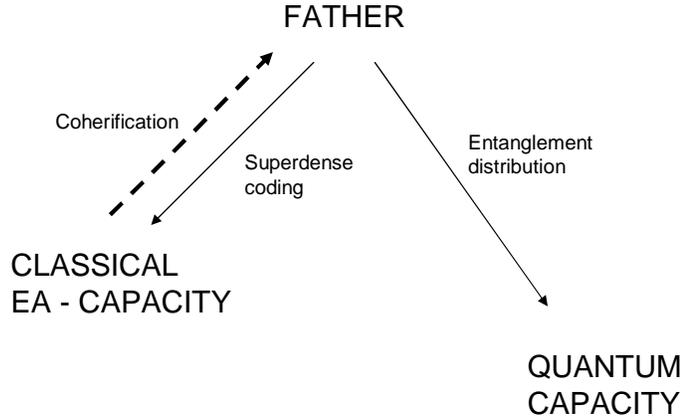}}}}
\caption{The male side of the family tree of quantum Shannon theory \cite{DHW03}.
%The so-called father protocol concerns sending quanutm information
%over a quantum channel assisted by entanglement. The quanutm capacity
%regards the same setting but without entanglement assistence.
%It is obtained from the father protocol via entanglement distrbution.
%The classical
}
\label{famfig}
\end{figure}

The father side of the family is shown in Figure \ref{famfig}.
Quantum  capacity-achieving protocols can be obtained from the
father protocol by combining it with entanglement distribution.
In conjunction with superdense coding,
the father protocol gives rise to EA capacity-achieving protocols.
Moreover, the latter can be made coherent \cite{Har03, Devetak03, DHW03, DHW05}
to recover the father protocol.

Can we reproduce the family in the finite setting of coding theory?
Is it beneficial to do so?
In this paper we give an affirmative answer to these two questions.
We  develop a general theory of linear ``father'' codes or
entanglement-assisted quantum error-correcting (EAQEC)
codes. The first and only such code to date
was constructed by Bowen \cite{Bow02} from the $[[5,1,3]]$ QEC code.
EAQEC codes turn out to be a rather natural generalization of the usual stabilizer codes, equivalent to classical symplectic codes.
These codes need not be dual-containing: the degree to which they
are not measures the required amount of entanglement assistance.
Consequently, \emph{any} classical code can be made into a EAQEC code.
This provides a drastic simplification, allowing the classical theory
of error correction to be imported wholesale.

The paper is organized as follows. Section 1 provides background on
the Pauli group and symplectic algebra. It also reviews basic quantum strategies for
sending classical information. Section 2 defines EAQEC codes
%entanglement assisted quantum error correcting codes,
and determines the set of errors they can correct.
Section 3 generalizes the code construction method of \cite{CRSS97, CRSS98}
based on classical codes over $\bbF_4$. Section 4 regards the right branch
of Figure \ref{famfig}: constructing \emph{catalytic} QEC codes from EAQEC codes.
Section 5 regards the left branch of Figure \ref{famfig}: constructing
entanglement assisted codes for sending classical information 
(EACEC codes).
These are then made coherent \cite{DHW05}, 
providing an alternative construction of EAQEC codes.
Section 6 discusses bounds on the performance of EAQEC codes. Section 7
recovers Bowen's result in our framework. Section 8
updates the table of known codes from \cite{CRSS98}.
%gives concrete examples of EAQEC codes and rules for their  shortening, puncturing, etc.
We discuss  our results  in Section 9.

\section{Background}

In this section we review the properties of Pauli matrices, and
relate them to symplectic binary and quaternary vector spaces.
Our presentation follows Forney et al. \cite{FGG05} and Hamada \cite{Hamada02}.

\subsection{Single qubit Pauli group}

A \emph{qubit} is a quantum system corresponding to a two
dimensional complex Hilbert space $\cH$.
Fixing a basis for $\cH$, the set $\Pi$ of \emph{Pauli matrices}
is defined as
$$
I = \left[\begin{array}{cc} 1 & 0 \\ 0 & 1 \end{array}\right],  \quad
X = \left[\begin{array}{cc} 0 & 1 \\ 1 & 0 \end{array}\right],  \quad
Y = \left[\begin{array}{cc} 0 & -i \\ i & 0 \end{array}\right],  \quad
Z = \left[\begin{array}{cc} 1 & 0 \\ 0 & -1 \end{array}\right].
$$
The Pauli matrices are Hermitian unitary matrices with eigenvalues belonging to
the set $\{1, -1 \}$.
The multiplication table of these matrices is given by:
$$
\begin{array}{|c|cccc|}  \hline
\times  & I & X & Y & Z \\  \hline
I & I & X & Y & Z \\
X & X & I & iZ & -iY \\
Y & Y & -iZ & I & iX \\
Z & Z & iY & -iX & I \\ \hline
\end{array}
$$
Observe that the Pauli matrices either commute or anticommute.
Let $[A] = \{\beta A \mid  \beta \in \bbC,
|\beta| = 1\}$ be the equivalence class of matrices equal to  $A$
up to a phase factor.\footnote{It makes good physical sense to 
neglect this overall phase, which has no observable consequence.} 
Then the set $[\Pi] = \{ [I], [X], [Y], [Z] \}$
is readily seen to form a commutative group under the multiplication
operation defined by $[A] [B] = [AB]$. It is called the Pauli group.

We are interested in relating the Pauli group to
the additive group $(\bbZ_2)^2 = \{00, 01, 10, 11\}$ of binary
words of length $2$ described by the table:
$$
\begin{array}{|c|cccc|} \hline
+  & 00 & 01 & 11 & 10 \\  \hline
00 & 00 & 01 & 11 & 10 \\
01 & 01 & 00 & 10 & 11 \\
11 & 11 & 10 & 00 & 01 \\
10 & 10 & 11 & 01 & 00 \\ \hline
\end{array}
$$
%Alternatively, the projective Pauli group is isomorphic to the
%additive group of the quaternary field $\bbF_4 = \{0, 1, \omega,
%\overline{\omega}\}$, whose addition table is
%$$
%\begin{array}{|c|cccc|}  \hline
%+  & 0 & \omega & 1 & \overline{\omega} \\  \hline
%0 & 0 & \omega & 1 & \overline{\omega} \\
%\omega & \omega & 0 & \overline{\omega} & 1 \\
%1 & 1 & \overline{\omega} & 0 & \omega \\
%\overline{\omega} & \overline{\omega} & 1 & \omega & 0 \\ \hline
%\end{array}
%$$
%In short, $0 + a = a$, $a + a = 0$ (so subtraction is the same as
%addition), and $1 + \omega + \overline{\omega} = 0$.
This group is also a two-dimensional vector space over the field
$\bbZ_2$. A bilinear form can be defined over this vector space, called the
\emph{symplectic form} or \emph{symplectic product}\footnote{Strictly
speaking it is not an inner product.}  $\odot: (\bbZ_2)^2 \times (\bbZ_2)^2
\rightarrow \bbZ_2$, given by the table
$$
\begin{array}{|c|cccc|} \hline
\odot  & 00 & 01 & 11 & 10 \\  \hline
00 & 0 & 0 & 0 & 0 \\
01 & 0 & 0 & 1 & 1 \\
11 & 0 & 1 & 0 & 1 \\
10 & 0 & 1 & 1 & 0 \\ \hline
\end{array}
$$
In what follows we will often write elements of $(\bbZ_2)^2$ as $u = (z|x)$, with $z,x \in \bbZ_2$.
For instance, $01$ becomes $(0|1)$.
For $u = (z|x), v = (z'|x') \in (\bbZ_2)^2$ the symplectic product is equivalently defined
by
$$
u \odot v = z x' - z' x.
$$
Define the map $N: (\bbZ_2)^2  \to  \Pi$ by the following table:
$$
\begin{array}{|c|c|} \hline
(\bbZ_2)^2 & \Pi\\  \hline
00 & I \\
01 & X \\
11 & Y \\
10 & Z \\ \hline
\end{array}
$$
This map is defined in such a way that $N_{(z|x)}$ and $Z^{z} X^{x}$
are equal up to a phase factor, i.e.
$$[N_{(z|x)}] = [ Z^{z} X^{x} ].$$
We make two key observations
\begin{enumerate}
\item  The map $[N] :  (\bbZ_2)^2  \to  [\Pi]$ induced by $N$
%defined by $ [N] = [ \, ] \circ N$
is an isomorphism:
$$
[N_u] [N_v] = [N_{u + v}].
$$
\item The commutation relations of the Pauli matrices are captured by the
symplectic product
$$
N_u N_v = (-1)^{u \odot v} N_v N_u.
$$
\end{enumerate}
Both properties are readily verified from the tables.

\subsection{Multi-qubit Pauli group}

Consider an $n$-qubit system corresponding to the
tensor product Hilbert space $\cH^{\otimes n}$.
Define an $n$-qubit Pauli matrix $\bA$ to be of the form
$\bA = A_1\otimes A_2\otimes \cdots\otimes A_n$, where $A_j \in \Pi$.
The set of all $4^n$ $n$-qubit Pauli matrices is denoted by $\Pi^n$.
The product of elements of $\Pi^n$ is an element of  $\Pi^n$ up to a
phase factor. Define as before the equivalence class
$[\bA] = \{\beta \bA \mid \beta \in \bbC, |\beta| = 1\}$.
Then
$$
[\bA] [\bB] = [A_1 B_1]\otimes[A_2 B_2]\otimes \cdots
\otimes[A_n B_n] = [\bA \bB].
$$
Thus the set $[\Pi^n] = \{ [\bA]: \bA \in \Pi^n \}$ is
a commutative multiplicative group.

Now consider the group/vector space $(\bbZ_2)^{2n}$ of binary vectors
of length $2n$.
Its elements may be written as $\bu = (\bz| \bx)$,
$\bz = z_1 \dots z_n \in (\bbZ_2)^n$,
$\bx = x_1 \dots x_n \in (\bbZ_2)^n$.
We shall think of $\bu$, $\bz$ and $\bx$ as row vectors.
The symplectic product of $\bu = (\bz|\bx)$ and $\bv = (\bz'|\bx')$
is given by
$$
\bu \odot \bv^{\T} = \bz \, {\bx'}^{\T} - \bz' \, \bx^{\T}.
$$
The right hand side are binary inner products and $T$ denotes the transpose.
This should be thought of as a kind of matrix multiplication of a row vector and a column vector.
We use $\bu \odot \bv^{\T}$ rather than the more standard $\bu \, \bv^\T$ to emphasize that the symplectic form 
is used rather than the binary inner product.
Equivalently,
$$
\bu \odot \bv^\T = \sum_i u_i \odot v_i
$$
where $u_i = (z_i|x_i), v_i = (z'_i|x'_i)$ and this sum represents Boolean addition.
Observe that $ {\bu} \odot {\bu}^\T = 0$, i.e., every vector is
``orthogonal'' to itself.

The map $N: (\bbZ_2)^{2n} \rightarrow \Pi^n $ is now defined as
$$
N_\bu = N_{u_1} \otimes \dots \otimes  N_{u_n}.
$$
Writing
$$
X^{\bx} = X^{x_1} \otimes  \dots  \otimes  X^{x_n},
$$
$$
Z^{\bz} = Z^{z_1} \otimes  \dots  \otimes  Z^{z_n},
$$
as in the single qubit case, we have
$$[N_{(\bz|\bx)}] = [ Z^{\bz} X^{\bx} ].$$

The two observations made for the single qubit case also hold:
\begin{enumerate}
\item  The map $[N] :  (\bbZ_2)^{2n}  \to  [\Pi^n]$ induced by $N$
%defined by $ [N] = [ \, ] \circ N$
is an isomorphism:
\be
[N_\bu] [N_\bv] = [N_{\bu + \bv}].
\label{eq:primi}
\ee
Consequently, if $\{ \bu_1, \dots, \bu_m \}$ is a
linearly independent set then the elements of the Pauli group subset
$\{ [N_{\bu_1}], \dots, [N_{\bu_m}] \}$ are independent
in the sense that no element can be written as a product of others.

\item The commutation relations of the $n$-qubit Pauli matrices are captured by the
symplectic product
\be
N_\bu N_\bv = (-1)^{\bu \odot \bv^\T} N_\bv N_\bu.
\label{eq:seconda}
\ee
\end{enumerate}

\medskip
%{\tt mention generators of Pauli subgroup vs bases for
%subspace of $(\bbZ_2)^{2n}$
%and how $[]$ has no effect on eigenspaces and measurements.}

\subsection{Properties of the symplectic form}
\label{sympo}

In this subsection we present two results which will play a major
role in the construction of EAQEC codes.
Together they will enable us to conclude that any independent subset
of the $n$-qubit Pauli group can be transformed via a unitary operation
into a canonical set whose elements act nontrivially only on single qubits.

A subspace $V$ of $(\bbZ_2)^{2n}$ is called \emph{symplectic} \cite{Silva01} if there 
is no $\bv \in V$ such that
\be
\bv \odot \bu^\T = 0,  \,\,\, \forall \bu \in V.
\label{eq:degen}
\ee
$(\bbZ_2)^{2n}$ is itself a symplectic subspace.
Consider the standard basis for $(\bbZ_2)^{2n}$,
consisting of $\bg_i = (\bee_i |\bo)$ and
$\bh_i = (\bo|\bee_i)$ for $i = 1, \dots ,n$,
where $\bee_i = (0,\dots, 0,1,0, \dots, 0)$ [$1$ in the $i$th position] are the standard
basis vectors of $(\bbZ_2)^n$. Observe that
\begin{eqnarray}
\bg_i \odot \bg_j^\T = 0, &  {\rm for \,\,  all \,\, } i,j \\
\bh_i \odot \bh_j^\T = 0, &  {\rm for \,\, all \,\, } i,j \\
\bg_i \odot \bh_j^\T = 0, &  {\rm for \,\,  all \,\, } i\neq j \\
\bg_i \odot \bh_i^\T = 1, &  {\rm for \,\,  all \,\, } i.
\end{eqnarray}
Thus, the basis vectors come in $n$ \emph{hyperbolic pairs} $(\bg_i, \bh_i )$
such that only the symplectic product between hyperbolic partners is nonzero.
The matrix $J = [\bg_i \odot \bh_j^\T]$ defining the symplectic product with respect
to this basis is given by
\be
\label{eq:syprma}
J =
\left(\begin{array}{cc}
0_{n \times n} & I_{n \times n} \\
 I_{n \times n} & 0_{n \times n}
\end{array} \right),
\ee
where $I_{n \times n}$ and $0_{n \times n}$ are the $n \times n$ identity
and zero matrices, respectively. A basis for $(\bbZ_2)^{2n}$ whose symplectic
product matrix $J$ is given by (\ref{eq:syprma}) is called a \emph{symplectic
basis}.
In the Pauli picture, the hyperbolic pairs $(\bg_i, \bh_i)$ correspond
to $(Z^{\bee_i},X^{\bee_i})$ -- the anticommuting $Z$ and $X$ Pauli matrices acting
on the $i$th qubit.

In contrast, a subspace $V$ of $(\bbZ_2)^{2n}$ is called \emph{isotropic}
if 
%$\odot|_{V} \equiv 0$. In other words, 
(\ref{eq:degen}) holds for \emph{all} $\bv \in V$.
%This gives a sense in which symplectic an isotropic are opposites.
The largest isotropic subspace of $(\bbZ_2)^{2n}$ is $n$-dimensional.
The span of  the $\bg_i$, $i = 1, \dots ,n$, is
an example of a subspace saturating this bound.

A general subspace of $(\bbZ_2)^{2n}$ is neither symplectic nor isotropic.
The following theorem, stated in \cite{Silva01}
and rediscovered in Pauli language in \cite{FCY+04},
says that an arbitrary subspace $V$ can
be decomposed as a direct sum of a symplectic part and an
isotropic part. 
We give an independent proof here.

\begin{theorem}
\label{thm1}
Let $V$ be an $m$-dimensional subspace of $(\bbZ_2)^{2n}$.
Then there exists a symplectic basis of $(\bbZ_2)^{2n}$
consisting of hyperbolic pairs $(\bu_i, \bv_i)$,
$i = 1, \dots ,n$, such that
$\{\bu_1, \dots, \bu_{c+ \ell}, \bv_1, \dots, \bv_{c}\}$ is a basis for $V$,
 for some $c, \ell \geq 0$ with $2c + \ell = m$.
%\begin{eqnarray}
%\bu_i \odot \bu_j^\T = 0, &  {\rm for \,\,  all \,\, } i,j \\
%\bv_i \odot \bv_j^\T = 0, &  {\rm for \,\, all \,\, } i,j \\
%\bu_i \odot \bv_j^\T = 0, &  {\rm for \,\,  all \,\, } i\neq j \\
%\bu_i \odot \bv_i^\T = 1, &  {\rm for \,\,  all \,\, } i.
%\end{eqnarray}

Equivalently,
$$
V =  \symp V \oplus \iso V
$$
where
$\symp V  = \spann { \bu_1, \dots,  \bu_c, \bv_1, \dots, \bv_{c} }$
is symplectic and $\iso V  = \spann { \bu_{c + 1}, \dots, \bu_{c + \ell} }$
is isotropic.
\end{theorem}

\noindent {\bf Remark}\,\, It is readily seen that the space $\iso{V}$ is unique,
given $V$. In contrast,  $\symp V$ is not. For instance, replacing
$\bv_1$ by $\bv_1' = \bv_1 +  \bu_{c + 1}$ in the above definition of $\symp V$
does not change its symplectic property.

\medskip

\begin{proof}
Pick an arbitrary basis $\{\bw_1, \dots, \bw_m \}$ for $V$ and
extend it to a basis $\{\bw_1, \dots, \bw_{2n} \}$ for $(\bbZ_2)^{2n}$.
We will describe an algorithm which yields the basis from the
statement of the theorem.

The procedure consists of $n$ rounds. In each round a new hyperbolic pair $(\bu_i, \bv_i)$
is generated; the index $i$ is added to the set $\cU$ $(\cV)$
if  $\bu_i \in V$  ($\bv_i \in V$).

Initially set $i =1$, $m' = m$, and $\cU = \cV = \emptyset$.
 % $V' = V$, $W = (\bbZ_2)^{2n}$.
 The $i$th round reads as follows.
\begin{enumerate}
\item We start with  vectors $\bw_1, \dots, \bw_{2(n-i+1)}$,
and $\bu_1, \dots \bu_{i-1},\bv_1, \dots \bv_{i-1}$,
%$\bw_1, \dots, \bw_{2(n-i+1)}$ are a basis for $W$, and
such that
\begin{enumerate}
\item $\bw_1, \dots, \bw_{2(n-i+1)},$
$\bu_1, \dots \bu_{i-1},$$\bv_1, \dots \bv_{i-1}$
is a basis for $(\bbZ_2)^{2n}$,
\item each of $\bu_1, \dots \bu_{i-1},\bv_1, \dots \bv_{i-1}$
has vanishing symplectic product with each of $\bw_1, \dots, \bw_{2(n-i+1)}$,
\item $V = {\rm span}\{ \bw_j: 1 \leq j \leq m'  \}
\oplus {\rm span}\{ \bu_j: j \in \cU \} \oplus {\rm span}\{ \bv_j: j \in \cV \}$.
\end{enumerate}
These conditions are  satisfied for $i=1$.

\item Define $\bu_i = \bw_1$.  If $m' \geq 1$ then and add $i$ to $\cU$.
%Observe that for each $\bu \in (\bbZ_2)^{2n}$ there
%exists some $v \in (\bbZ_2)^{2n}$ such that ${\bu} \odot {\bv} = 1$.
Let $j\geq 2$ be the smallest index for which
$\bw_1 \odot \bw_j^\T = 1$. Such a $j$ exists because of (a), (b) and
the fact that there exists a $\bw \in (\bbZ_2)^{2n}$
such that ${\bu_i} \odot {\bw}^\T = 1$.

Set $\bv_i = \bw_j$.
\item If $j \leq m'$:

This means that there is a  hyperbolic  partner of $\bu_i$ in $V$.
 Add $i$ to $\cV$;
swap $\bw_j$ with $\bw_2$;
for $k = 3, \dots, 2(n-i+1)$ %and excluding $k = j$
perform
$$
\bw'_{k-2} := \bw_k - (\bv_i \odot \bw_k^\T) \bu_i -  (\bu_i \odot \bw_k^\T) \bv_i,
$$
so that
\be
\bw'_{k-2} \odot  \bu_i^\T = \bw'_{k-2} \odot  \bv_i^\T = 0 ;
\label{upravo1}
\ee
set $m' := m' - 2$.

If $j > m'$:

This means that there is no hyperbolic partner of $\bu_i$ in $V$.
Swap $\bw_j$ with $\bw_{2(n-i+1)}$; for $k = 2, \dots,2(n-i)+1$ perform
$$
\bw'_{k-1} := \bw_k - (\bv_i \odot \bw_k^\T) \bu_i -  (\bu_i \odot \bw_k^\T) \bv_i,
$$
so that
\be
\bw'_{k-1} \odot  \bu_i^\T = \bw'_{k-1} \odot  \bv_i^\T = 0 ;
\label{upravo2}
\ee
if $m' \geq 1$ then set $m' := m' - 1$.
\item Let $\bw_k := \bw'_k$ for $1 \leq k \leq 2(n - i)$.
We need to show that the conditions from
item 1 are satisfied for the next round ($i: = i + 1$).
Condition (a) holds because
 $\{ \bu_i, \bv_i, \bw'_1, \dots \bw'_{2(n-i)}\}$
are related to the
old $\{\bw_1, \dots \bw_{2(n-i+1)}\}$
by an invertible linear transformation.
Condition (b) follows from (\ref{upravo1}) and (\ref{upravo2}).
Regarding condition (c), if $m' = 0$ then it holds
because $\cU$ and $\cV$ did not change from the previous round.
Otherwise, consider the two cases in item 3.
If $j \leq m'$ then $\{ \bu_i, \bv_i, \bw'_1, \dots \bw'_{m'-2}\}$
are related to the old $\{\bw_1, \dots \bw_{m'}\}$
by an invertible linear transformation.
If $j > m'$ then $\{ \bu_i, \bw'_1, \dots \bw'_{m'-1}\}$
are related to the old $\{\bw_1, \dots \bw_{m'}\}$
by an invertible linear transformation
(the  $(\bu_i \odot \bw_k^\T) \bv_i$ terms vanish for $1 \leq k \leq m'$
because there is no hyperbolic partner of $\bu_i$ in $V$).

\end{enumerate}
$0 \leq m' \leq 2(n-i)$ at the end of the $i$th round.
Thus $m' = 0$ after $n$ rounds and hence
$V =  {\rm span}\{ \bu_j: j \in \cU \} \oplus {\rm span}\{ \bv_j: j \in \cV \}$.
The theorem follows by suitably reordering the $(\bu_j,\bv_j)$.
\end{proof}

\medskip

A \emph{symplectomorphism} $\Upsilon:(\bbZ_2)^{2n} \rightarrow
(\bbZ_2)^{2n} $  is a linear isomorphism which
preserves the symplectic form, namely
\be
\Upsilon (\bu) \odot
\Upsilon (\bv)^\T =  \bu \odot \bv^\T.
\label{smor}
\ee
%Given a symplectic basis $\cB = \{(\bu_i, \bv_i): i = 1, \dots, n \}$ the
%map $\Upsilon$ defined by
%\begin{eqnarray}
%\Upsilon (\bg_i) &= \bu_i \label{eq:gu} \\
%\Upsilon (\bh_i) &= \bv_i \label{eq:hv}
%\end{eqnarray}
%is a  symplectomorphism which maps
%the standard basis  $\{ (\bg_i, \bh_i): i = 1, \dots, n \}$
%to $\cB$.

The following theorem relates symplectomorphisms
on $(\bbZ_2)^{2n}$ to unitary maps on $\cH^{\otimes n}$.
It appears, for instance, in \cite{BFG05}. For completeness, 
we give an independent proof here.

\begin{theorem} \label{thm2}
For any symplectomorphism  $\Upsilon$ on $(\bbZ_2)^{2n}$
 there exists a unitary map
$U_\Upsilon$ on $\cH^{\otimes n}$ such that for all
$\bu \in (\bbZ_2)^{2n}$,
$$
[N_{\Upsilon (\bu)}] = [ U_\Upsilon N_{\bu} U_\Upsilon^{-1} ].
$$
\end{theorem}

\noindent {\bf Remark.}\, The unitary map $U_\Upsilon$ may be viewed as a
map on $[\Pi]$ given by  $[\bA] \mapsto [U_\Upsilon \bA U_\Upsilon^{-1}]$.
The theorem says that the following diagram commutes
$$
\begin{CD}
(\bbZ_2)^{2n}   @>{{\Upsilon}}>> (\bbZ_2)^{2n}   \\
@V{{[N]}}VV      @VV{{[N]}}V \\
[\Pi]   @>U_\Upsilon>> [\Pi]
\end{CD}
$$

\begin{proof}
Consider the standard basis
$\bg_i = (\bee_i |\bo)$, $\bh_i = (\bo|\bee_i)$.
Define the unique (up to a phase factor) state $\ket{\bo}$ on $\cH^{\otimes n}$
 to be the simultaneous $+1$
eigenstate of the commuting operators $N_{\bg_j}$, $j = 1, \dots, n$.
Define an orthonormal basis
$\{ \ket{\bb}: \bb = b_1 \dots b_n \in (\bbZ_2)^{n} \}$
for $\cH^{\otimes n}$ by
$$
\ket{\bb} = N_{\sum_i b_i \bh_i} \ket{\bo}.
$$
The orthonormality follows from the observation that
$\ket{\bb}$ is a simultaneous eigenstate of $N_{\bg_j}$, $j = 1, \dots, n$
with respective eigenvalues $(-1)^{b_j}$:
\be
\begin{split}
N_{\bg_j} \ket{\bb} & = N_{\bg_j}  N_{\sum_i b_i \bh_i} \ket{\bo}\\
& = (-1)^{b_j}  N_{\sum_i b_i \bh_i}  N_{\bg_j}  \ket{\bo} \\
& = (-1)^{b_j}  N_{\sum_i b_i \bh_i}  \ket{\bo} \\
& = (-1)^{b_j} \ket{\bb}.
\end{split}
\ee
The second line is an application of (\ref{eq:seconda}).

Define $\tilde{\bg}_i: =  \Upsilon (\bg_i)$. %Repeat the above.
We repeat the above construction for this new basis.
Define the unique (up to a phase factor) state ${\ket{\tilde{\bo}}}$ to be the simultaneous $+1$
eigenstate of the commuting operators
$N_{\tilde{\bg}_i}$, $i = 1, \dots, n$.
Define an orthonormal basis $\{ {\ket{\tilde{\bb}}} \}$ by
\be
\ket{\tilde{\bb}} = N_{ \sum_i b_i \tilde{\bh}_i} \ket{\tilde{\bo}}.
\label{ono}
\ee
%We need to show that
%$$
%\tilde{ N_\bu \ket{\bb}} = N_{\tilde{\bu}} \ket{\tilde{\bb}}.
%$$
Defining
$\bu =  \sum_i z_i \bg_i + x_i \bh_i$,
$\tilde{\bu} = \sum_i z_i \tilde{\bg}_i + x_i \tilde{\bh}_i $
and $\bx = x_1 \dots x_n$, we have
\be
\label{eq:ludak1}
\begin{split}
N_{\tilde{\bu}} \ket{\tilde{\bb}}
& =  N_{\tilde{\bu}} N_{ \sum_i  b_i \tilde{\bh}_i} {\ket{\tilde{\bo}}} \\
& = (-1)^{ \tilde{\bu} \odot  (\sum_i  b_i \tilde{\bh}_i)^\T }
 N_{ \sum_i  b_i \tilde{\bh}_i} N_{\tilde{\bu}} {\ket{\tilde{\bo}}} \\
& = (-1)^{ \tilde{\bu} \odot  (\sum_i  b_i \tilde{\bh}_i)^\T }
e^{i \theta(\tilde{\bu})}
 N_{ \sum_i  b_i \tilde{\bh}_i}
 N_{\sum_i  x_i \tilde{\bh}_i}  N_{\sum_i  z_i \tilde{\bg}_i}
 {\ket{\tilde{\bo}}} \\
&= (-1)^{ \tilde{\bu} \odot  (\sum_i  b_i \tilde{\bh}_i)^\T }
e^{i \theta(\tilde{\bu})}
 N_{ \sum_i ( b_i + x_i) \tilde{\bh}_i}
 {\ket{\tilde{\bo}}} \\
&= (-1)^{ \tilde{\bu} \odot  (\sum_i  b_i \tilde{\bh}_i)^\T }
e^{i \theta(\tilde{\bu})}
 {\ket{\tilde{\bb + \bx }}} \\
&= (-1)^{ {\bu} \odot  (\sum_i  b_i {\bh}_i)^\T }
e^{i \theta(\tilde{\bu})}
 {\ket{\tilde{\bb + \bx }}},
\end{split}
\ee
where $\theta(\tilde{\bu})$ is a phase factor which is
independent of $\bb$.
The first equality follows from (\ref{ono}),
the second  from (\ref{eq:seconda}), the
third from (\ref{eq:primi}), the fourth from the
definition of $\ket{\tilde{\bo}}$ and the fact that
$X^{\bb} X^{\bx} = X^{\bb +\bx}$, the fifth from (\ref{ono}),
and the sixth from (\ref{smor}).
Similarly
\be
\label{eq:ludak2}
 N_\bu \ket{\bb} = (-1)^{{\bu} \odot  (\sum_i  b_i {\bh}_i)^\T }
 e^{i \varphi({\bu})}
 {\ket{{\bb + \bx }}},
\ee
where $\varphi(\bu)$ is a 
is a phase factor which is
independent of $\bb$.

Define $U_\Upsilon$ by the change of basis
$$
U_\Upsilon = \sum_\bb \ket{\tilde{\bb}}\bra{{\bb}}.
$$
Combining (\ref{eq:ludak1}) and (\ref{eq:ludak2})
gives for all $\ket{\bb}$
\be
\begin{split}
N_{ \Upsilon (\bu)} U_\Upsilon \ket{\bb} &=
 (-1)^{ {\bu} \odot  (\sum_i  b_i {\bh}_i)^\T }
e^{i \theta(\tilde{\bu})}   U_\Upsilon  \ket{{\bb + \bx }} \\
&= e^{i [\theta(\tilde{\bu}) -   \varphi({\bu})]}
 U_\Upsilon N_\bu \ket{\bb}.
\end{split}
\ee
Therefore $[N_{\Upsilon (\bu)}] = [ U_\Upsilon N_{\bu} U_\Upsilon^{-1} ]$.
\end{proof}

\subsection{Encoding classical information into quantum states}
\label{eci}
In this subsection we review two schemes for sending
classical information over quantum channels: elementary coding and
superdense coding. These will be used later in the context of quantum error correction
to convey information to the decoder
about which error happened.

In the first scheme,
Alice and Bob are connected by
a perfect qubit channel. Alice can send an arbitrary bit
$a \in \bbZ_2$ over the qubit channel in the following way:
\begin{itemize}
\item Alice locally prepares a state $\ket{0}$ in $\cH$.
This state is the $+1$ eigenstate of the $Z$ operator.
Based on her message $a$, she performs the encoding operation $X^{a}$,
producing the state $\ket{a}$. 
\item Alice sends the encoded state to Bob
through the  qubit channel.
\item Bob decodes by performing the von Neumann  measurement in the
$\{\ket{0}, \ket{1} \}$ basis.
As this is the unique eigenbasis of
the $Z$ operator, this is equivalently called ``measuring the $Z$
observable''.
\end{itemize}
We call this protocol ``elementary coding'' and
write it symbolically as a \emph{resource inequality} \cite{DHW05, DHW03, DW03a}
\footnote{In  \cite{DHW05} resource inequalities were used in
the asymptotic sense. Here they refer to finite protocols, and
are thus slightly abusing their original intent.}
$$
[q \rightarrow q] \geq [c \rightarrow c].
$$
Here $[q \rightarrow q]$ represents a perfect qubit channel
and $[c \rightarrow c]$ represents a perfect classical bit channel.
The inequality $\geq$ signifies that the resource on the left hand side can be
used in a protocol to simulate the resource on the right hand side.

Elementary coding immediately extends to $m$ qubits.
Alice prepares the simultaneous $+1$ eigenstate of
the $Z^{\bee_1}, \dots, Z^{\bee_m}$ operators $\ket{\bo}$, and encodes the message
$\ba \in (\bbZ_2)^m$ by applying
$X^{\ba}$.
Bob decodes by simultaneously
measuring the $Z^{\bee_1}, \dots, Z^{\bee_m}$ observables.
We could symbolically represent this protocol by
$$
m \,[q \rightarrow q] \geq m \,[c \rightarrow c].
$$
%It is not necessary for the initial state to be the simultaneous
%$+1$ eigenstate of $Z^{\bee_1}, \dots, Z^{\bee_m}$.
%Any of the $2^m$ initial states $\ket{\bff} = X^{\bff} \ket{\bo}$,
%$\bff= f_1 \dots f_m \in (\bbZ_2)^m$,
%works just as well. The state $\ket{\bff}$ is the $(-1)^{f_i}$ eigenstate
%of $Z^{\bee_i}$.
% The encoded state becomes $\ket{\bff + \ba}$, and
%$\bff$ must be subtracted from  measurement result, $\bff + \ba$,
%to correctly decode $\ba$.

\medskip

In the second scheme, Alice and Bob share the ebit state
\be
\label{eq:ebit}
\ket{\Phi} = \frac{1}{\sqrt{2}} (\ket{0} \otimes  \ket{0} +
\ket{1} \otimes \ket{1})
\ee
in addition to being connected by the qubit channel.
In (\ref{eq:ebit}) Alice's state is to the left and
Bob's is to the right of the $\otimes$ symbol.

The state $\ket{\Phi}$ is the simultaneous $(+1, +1)$ eigenstate of
the commuting operators  $Z \otimes Z$ and  $X \otimes X$.
Again, the operator to the left of the $\otimes$ symbol
acts on Alice's system and the operator to  the right of the $\otimes$ symbol
acts on Bob's system.
Alice can send a two-bit message $(a_1,a_2) \in (\bbZ_2)^2$
to Bob using ``superdense coding'' \cite{BW92}:
\begin{itemize}
\item Based on her message $(a_1, a_2)$, Alice performs the encoding
operation $Z^{a_1} X^{a_2}$ on her part of the state $\ket{\Phi}$.
\item Alice sends her part of the encoded state to Bob
through the perfect qubit channel.
\item Bob decodes by performing the von Neumann measurement in the
$\{(Z^{a_1} X^{a_2} \otimes I )\ket{\Phi}: (a_1, a_2) \in (\bbZ_2)^2 \}$ basis,
i.e., by simultaneously measuring the
$Z \otimes Z$ and  $X \otimes X$ observables.
\end{itemize}
The protocol is represented by the resource inequality
\be
[q \rightarrow q] + [q \, q] \geq 2 \, [c \rightarrow c],
\label{eq:sd}
\ee
where $[q \,q]$ now represents the shared ebit.
It can also  be extended to $m$ copies.
Alice and Bob share the state $\ket{\Phi}^{\otimes m}$
which is the simultaneous $+1$ eigenstate of
the $Z^{\bee_1} \otimes Z^{\bee_1}, \dots, Z^{\bee_m} \otimes Z^{\bee_m}$
and $X^{\bee_1} \otimes X^{\bee_1}, \dots, X^{\bee_m} \otimes X^{\bee_m}$
operators. Alice  encodes the message
$(\ba_1, \ba_2) \in (\bbZ_2)^{2m}$
by applying $X^{\ba_1} Z^{\ba_2}$.
Bob decodes by simultaneously
measuring the
 $Z^{\bee_1} \otimes Z^{\bee_1}, \dots, Z^{\bee_m} \otimes Z^{\bee_m}$
and $X^{\bee_1} \otimes X^{\bee_1}, \dots, X^{\bee_m} \otimes X^{\bee_m}$
 observables.
The corresponding resource inequality is
$$
m \, [q \rightarrow q] + m \,[q \, q] \geq 2m \, [c \rightarrow c].
$$

Superdense coding provides the simplest illustration of how entanglement can increase
the power of information processing.

\section{Entanglement-assisted quantum error correction}
\label{sec:EAQECcode}

In this section we introduce entanglement-assisted error correcting (EAEC) codes and prove our main result, 
Theorem 2.4, which gives sufficient error-correcting conditions.

\subsection{The model: discretization of errors}

\begin{figure}
\centerline{ {\scalebox{0.50}{\includegraphics{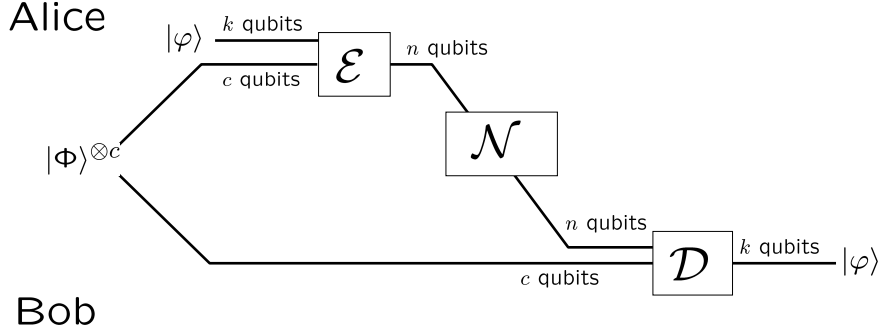}}}}
\caption{A generic entanglement assisted quantum code.}
\label{figfather}
\end{figure}

It is well known that for standard quantum error correction
(i.e., that unassisted by entanglement) it suffices to consider
errors from the Pauli group (see e.g. \cite{NC00}.) We will show this for entanglement-assisted quantum error
correction.
Denote by $\cL$ the space of linear operators defined on the
qubit Hilbert space $\cH$. We will often encounter
isometric operators $U: \cH^{\otimes n_1} \rightarrow  \cH^{\otimes n_2}$.
The corresponding \emph{superoperator}, or completely positive,
trace preserving (CPTP) map, is marked by a hat
$\hat{U}: \cL^{\otimes n_1} \rightarrow \cL^{\otimes n_2}$
and defined by
$$
\hat{U}(\rho) = U \rho U^\dagger.
$$
Observe that $\hat{U}$ is independent of any phases factors
multiplying $U$. Thus, for a  Pauli operator $N_\bu$,
$\hat{N}_\bu$ only depends on the equivalence class $[N_\bu]$.

Our communication scenario involves two spatially separated
parties, Alice and Bob, as depicted in Figure \ref{figfather}.
The resources at their disposal are
\begin{itemize}
\item a noisy channel
defined by a CPTP map
$\cN:  \cL^{\otimes n} \rightarrow \cL^{\otimes n}$
taking density operators on Alice's system to density operators
on Bob's system;
\item the $c$ ebit state $\ket{\Phi}^{\otimes c}$ shared
between Alice and Bob.
\end{itemize}
Alice wishes to send $k$ qubits  \emph{perfectly}
to Bob using the above resources.
An $[[n,k;c]]$ entanglement-assisted quantum error correcting (EAQEC) code
 consists of
\begin{itemize}
\item An encoding isometry $\cE = \hat{U}_{\rm enc}:
 \cL^{\otimes k} \otimes \cL^{\otimes c} \rightarrow \cL^{\otimes n}$
\item A decoding CPTP map
$\cD: \cL^{\otimes n} \otimes \cL^{\otimes c} \rightarrow \cL^{\otimes k}$
\end{itemize}
such that
$$
 \cD \circ \cN \circ \hat{U}_{\rm enc} \circ \hat{U}_{\rm app} = \id^{\otimes k},
$$
where ${U}_{\rm app}$ is the isometry which appends the state
$\ket{\Phi}^{\otimes c}$,
$$
U_{\rm app} \ket{\varphi} =  \ket{\varphi} \ket{\Phi}^{\otimes c},
$$
and $\id: \cL \rightarrow \cL$ is the identity map on a single qubit.
%This is shown in Figure \ref{eaqecc}.
The protocol thus uses up $c$ ebits of entanglement and generates
$k$ perfect qubit channels. We represent it by the resource inequality
(with a slight abuse of notation \cite{DHW05})
$$
\< \cN \>  + c \, [q \, q] \geq  k \, [q \rightarrow q].
$$
Even though a qubit channel is a strictly stronger
resource than its static analogue, an ebit of entanglement, the parameter $k-c$
is still a good (albeit pessimistic) measure of the net noiseless quantum resources
gained. It should be borne in mind that a negative value of $k-c$ still
refers to a non-trivial protocol.

To make contact with classical error correction it is
necessary to discretize the errors. This is done in two steps.
First, the CPTP map $\cN$ may be (non-uniquely) written in
terms of its Kraus representation
$$
\cN(\rho) = \sum_i A_i \rho A_i^\dagger.
$$
Second, each $A_i$ may be expanded in the Pauli operators
$$
A_i = \sum_{\bu \in  (\bbZ_2)^{2n}} \alpha_{i, \bu} N_\bu.
$$
Define the support of $\cN$ by
${\rm supp}(\cN) =
\{\bu \in (\bbZ_2)^{2n}: \exists i, \alpha_{i, \bu} \neq 0 \}$.
The following theorem allows us, absorbing ${U}_{\rm app}$ into ${U}_{\rm enc}$,
to replace the
continuous map $\cN$ by the
error set $S = {\rm supp}(\cN)$.

\begin{theorem}
 \label{discretization}
If $\cD \circ \hat{N}_\bu \circ \hat{U}_{\rm enc} = \id^{\otimes k}$
for all $\bu \in {\rm supp}(\cN)$, then
 $\cD \circ \cN \circ \hat{U}_{\rm enc} = \id^{\otimes k}$.
\end{theorem}

\begin{proof}
We may extend the map $\cD$ to its Stinespring dilation --
an isometric map $\hat{U}_{\rm dec}$  with a larger target Hilbert space
$\cL^{\otimes k} \otimes \cL'$, such that
$$
\cD = \tr_{\cL'} \circ \hat{U}_{\rm dec}.
$$
The premise of the theorem is equivalent to saying that
for all $\bu \in {\rm supp}(\cN)$ and all pure states
$\ket{\varphi}$ in $\cH^{\otimes n}$,
$$
U_{\rm dec}  N_\bu {U}_{\rm enc} \ket{\varphi} =  \ket{\varphi} \otimes \ket{\bu}
$$
for some pure state $\proj{\bu}$ on $\cL'$.
By linearity
$$
U_{\rm dec} A_i {U}_{\rm enc} \ket{\varphi} =
\ket{\varphi} \otimes \ket{i},
$$
with the unnormalized state $\ket{i} =  \sum_\bu \alpha_{i, \bu} \ket{\bu}$.
Furthermore,
\be
\begin{split}
(\hat{U}_{\rm dec} \circ \cN \circ
 \hat{U}_{\rm enc}) (\proj{\varphi})
&= U_{\rm dec}
\left( \sum_i A_i {U}_{\rm enc} \proj{\varphi} {U}_{\rm enc}^\dagger A_i^\dagger \right)
 U_{\rm dec}^\dagger\\
& =  \proj {\varphi} \otimes \sum_i \proj{i},
\end{split}
\ee
where the second subsystem corresponds to $\cL'$.
Tracing out the latter gives
$$
(\cD \circ \cN \circ \hat{U}_{\rm enc}) (\proj{\varphi}) = \proj {\varphi},
$$
concluding the proof.
\end{proof}

\subsection{The canonical error set and  syndrome coding}

By the results of the previous subsection, we are now interested
in EAQEC codes which correct a particular error set
$S \subset (\bbZ_2)^{2n}$. We first restrict attention
to a simple error set, which will turn out
to be generic due to the results of Section
\ref{sympo}.

Consider the following trivial encoding operation
$\hat{U}_{0}$ defined by
\be
U_0: \ket{\varphi} \ket{\Phi}^{\otimes c}  \mapsto
\ket{\varphi} \ket{\bo} \ket{\Phi}^{\otimes c}  .
\label{tren}
\ee
In other words, the register containing $\ket{\bo}$
(of size $\ell = n-k-c$ qubits) is appended to the
registers containing $\ket{\varphi}$ (of size
$k$ qubits) and $\ket{\Phi}^{\otimes c}$ (of
size $c$ qubits each for Alice and Bob).
What errors can she correct with such a simple-minded encoding?

\medskip

\begin{proposition}
%\noindent{\bf Claim.} \,
The code given by $U_0$ and a suitably defined decoding map
$\cD_0$ can correct the error set
\be
\label{sdef}
S_0 = \{ \left(  \bal(\ba, \ba_1, \ba_2), \bb, \ba_2
|
\bbe(\ba, \ba_1, \ba_2), \ba, \ba_1  \right):
\bb, \ba \in (\bbZ_2)^{\ell},
\ba_1, \ba_2 \in (\bbZ_2)^{c} \},
\ee
for any functions $\bal,\bbe: (\bbZ_2)^{\ell} \times (\bbZ_2)^{c} \times
(\bbZ_2)^{c} \rightarrow (\bbZ_2)^{k}$.
\end{proposition}

%\medskip

\begin{figure}
\centerline{ {\scalebox{0.50}{\includegraphics{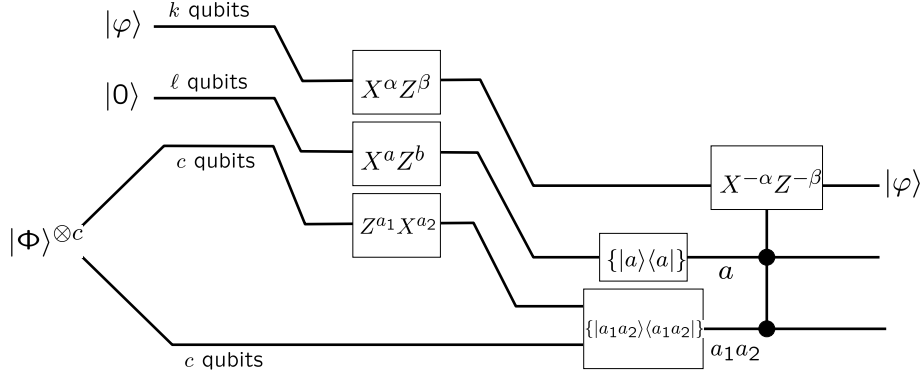}}}}
\caption{The canonical code.}
\label{figdumb}
\end{figure}

\begin{proof}
The protocol is shown in figure \ref{figdumb}.
After applying an error $N_\bu$ with
\be
\bu = ( \bal(\ba, \ba_1, \ba_2), \bb, \ba_2 |
\bbe(\ba, \ba_1, \ba_2) , \ba, \ba_1),
\label{boo}
\ee
the encoded state $ \ket{\varphi} \ket{\bo} \ket{\Phi}^{\otimes c}$
becomes (up to a phase factor)
\be
\begin{split}
&
 Z^{\bal(\ba, \ba_1, \ba_2)} X^{\bbe(\ba, \ba_1, \ba_2)}  \ket{\varphi}
\otimes
X^{\ba} Z^\bb \ket{\bo}
\otimes (Z^{\ba_1} X^{\ba_2} \otimes I) \ket{\Phi}^{\otimes c}
 \\
&=  Z^{\bal(\ba, \ba_1, \ba_2)}
X^{\bbe(\ba, \ba_1, \ba_2)}  \ket{\varphi} \otimes
\ket{\ba}
\otimes  \ket{\ba_1, \ba_2},
\end{split}
\label{zadnja}
\ee
where $\ket{\ba} =   X^{\ba} \ket{\bo}$
and $\ket{\ba_1, \ba_2} =
(Z^{\ba_1} X^{\ba_2} \otimes I) \ket{\Phi}^{\otimes c}$.
As the vector $(\ba, \ba_1, \ba_2, \bb)^\T$ completely specifies
the error $\bu$, it is called the \emph{error syndrome}.
The state (\ref{zadnja}) only depends on the \emph{reduced syndrome}
$\br = (\ba, \ba_1, \ba_2)^\T$. In effect, $\ba$ and $(\ba_1, \ba_2)$ have been
encoded using plain and superdense coding, respectively.
Bob, who holds the entire state (\ref{zadnja}), may
identify the reduced syndrome using the results of section
\ref{eci}.
Bob simultaneous measures the
$Z^{\bee_1}, \dots, Z^{\bee_m}$ observables
to decode $\ba$,
the $Z^{\bee_1} \otimes Z^{\bee_1}, \dots, Z^{\bee_m} \otimes Z^{\bee_m}$
observables to decode $\ba_1$, and the
$X^{\bee_1} \otimes X^{\bee_1}, \dots, X^{\bee_m} \otimes X^{\bee_m}$
observables to decode $\ba_2$. He then performs
$Z^{-\bal(\ba, \ba_1, \ba_2)} X^{-\bbe(\ba, \ba_1, \ba_2)}$
on the remaining $k$ qubit system, leaving it in the state
$\ket{\varphi}$.

Since the goal is the transmission of quantum
information, no actual measurement is necessary.
Instead, Bob can perform the CPTP map $\cD_0$
consisting of the controlled  unitary
$$
U_{0 \, \rm{dec}} = \sum_{\ba, \ba_1, \ba_2}
Z^{-\bal(\ba, \ba_1, \ba_2)} X^{-\bbe(\ba, \ba_1, \ba_2)}
   \otimes
  \proj{\ba}
\otimes  \proj{\ba_1, \ba_2},
$$
followed by discarding the last two subsystems.
\end{proof}

\medskip

The above code is  \emph{degenerate} with respect to
the error set $S$, which means that the error can be corrected
without knowing the full error syndrome.
% This is commonly interepreted as
%``the error can be corrected without knowing it completely''.
%An extreme example is that of decoherence-free subspaces.
%For error detection perform an incomplete binary measurement
%of the zeroness of the reduced syndrome.

We can characterize our code in terms of
the \emph{parity check matrix} $\H$ given by
\be
\label{uno1}
\H = \left(\begin{array}{c}
\H_I \\
\H_S
\end{array} \right),
\ee
\be \label{uno2}
\H_I =
\left(\begin{array}{ccc|ccc}
\bo_{\ell \times k} & \bI_{\ell \times \ell} & \bo_{\ell \times c}
& \bo_{\ell \times k} & \bo_{\ell \times \ell} & \bo_{\ell \times c}
\end{array} \right),
\ee
\be \label{uno3}
\H_S = \left(\begin{array}{ccc|ccc}
 \bo_{c \times k} & \bo_{c \times \ell} & \bI_{c \times c}  &
 \bo_{c \times k} & \bo_{c \times \ell} & \bo_{c \times c}  \\
\bo_{c \times k} & \bo_{c \times \ell} & \bo_{c \times c}
 & \bo_{c \times k} & \bo_{c \times \ell} & \bI_{c \times c}
\end{array} \right),
\ee
with $\ell = n-k-c$.

The vector space $\rs{\H}$ decomposes into a direct sum of
the isotropic subspace $\rs{\H_I}$ and symplectic subspace $\rs{\H_S}$,
as in Theorem \ref{thm1}.
 Define the \emph{symplectic code} corresponding to
$\H$ by
$$
C_0 = \rs{\H}^\perp
$$
where
$$
V^\perp = \{\bw: \bw \odot \bu^\T = 0, \,\, \forall \bu \in V\}.
$$
Note that $(V^\perp)^\perp = V$. %We will often abbreviate $C(\H)$ to $C_0$.
Then $C_0^\perp = \rs{\H}$, $\iso{C_0^\perp} = \rs{\H_I}$ and
 $\symp{C_0^\perp} = \rs{\H_S}$.

The number of ebits used in the code is
$$
c = \frac{1}{2}\dim \rs{\H_S}
$$
and the number of
encoded qubits is
$$
k = n - \dim \rs{\H_I} - {\small \frac{1}{2}} \dim \rs{\H_S}.
$$
The parameter $k - c$ which is the
%As we will discuss later in the context of catalysis,
%a particularly important parameter is the
number of encoded qubits minus the number of ebits used
is independent of the symplectic structure of $\H$:
$$
{k}-c = n - \dim \rs{\H}.
$$

The error set $S_0$ can be described in terms of $\H$:
\begin{proposition}
\label{ecc}
The set $S_0$ of errors correctable by the code $\cE_0$ is
such that, if $\bu, \bu' \in S_0$ and $\bu \neq \bu'$,
then either
\begin{enumerate} [i)]
\item $\bu - \bu' \not\in C_0$ (equivalently:
 $\H \odot (\bu - \bu')^\T \neq \bo^\T$), or 
\item $\bu - \bu' \in \iso{C_0^\perp}$
(equivalently: $\bu - \bu' \in \rs{\H_I}$)
%$\bu - \bu' \in \iso{C_0^\perp}$.
\end{enumerate}
\end{proposition}

\begin{proof}
If $\bu$ is given by (\ref{boo}) then $\H \odot \bu^\T = \br = (\ba, \ba_1, \ba_2)^\T$,
the reduced error syndrome.
By definition (\ref{sdef}), two distinct elements of $S_0$ either have different reduced syndromes
$(\ba, \ba_1, \ba_2)$ (condition 1) or they differ by a vector
of the form $( \bo, \bb, \bo |  \bo,  \bo,  \bo)$ (condition 2).
Observe that condition 1 is analogous to the usual error correcting
condition for classical codes \cite{FJM77}.
\end{proof}

\medskip

The parity check matrix $\H$ also specifies the encoding and decoding
operations.
The space $\cH^{\otimes k}$ is encoded into the \emph{codespace} defined by
$$
\cC_0 = \{ U_0 \ket{\varphi} \ket{\Phi}^{\otimes c}  :
\ket{\varphi} \in \cH^{\otimes k}\}.
$$
%where $U_{\cE_0}$ is the unique Kraus element of $\cE_0$.
%into which $\cH^{\otimes k}$ is encoded is the
It is not hard to see that the codespace is the
simultaneous $+1$ eigenspace of the commuting operators:
\begin{enumerate}
\item $I \otimes Z^{\bee_i} \otimes I \otimes I, \,\,i= 1, \dots, \ell$;
\item $I \otimes I \otimes Z^{\bee_j} \otimes Z^{\bee_j}, \,\, j= 1, \dots, c$;
\item $I \otimes I \otimes X^{\bee_j} \otimes X^{\bee_j} , \,\, j= 1, \dots, c$.
\end{enumerate}
Above, the first three operators act on Alice's qubits and the fourth on Bob's.
Define the matrix
\be B =
 \left(\begin{array}{c|c}
 \bo_{\ell \times c}  & \bo_{\ell \times c} \\
 \bI_{c \times c}  & \bo_{c \times c} \\
 \bo_{c \times c}  & \bI_{c \times c}
\end{array} \right).
\label{eq:beba}
\ee
Define the \emph{augmented} parity check matrix
\be
\H_{\rm aug} = (\H, B)  =
 \left(\begin{array}{cccc|cccc}
\bo_{\ell \times k} & \bI_{\ell \times \ell} & \bo_{\ell \times c}  & \bo_{\ell \times c}
& \bo_{\ell \times k} & \bo_{\ell \times \ell} & \bo_{\ell \times c} & \bo_{\ell \times c}
\\
 \bo_{c \times k} & \bo_{c \times \ell} & \bI_{c \times c} & \bI_{c \times c}
 &
 \bo_{c \times k} & \bo_{c \times \ell} & \bo_{c \times c} & \bo_{c \times c} \\
\bo_{c \times k} & \bo_{c \times \ell} & \bo_{c \times c}  & \bo_{c \times c}
 & \bo_{c \times k} & \bo_{c \times \ell} & \bI_{c \times c} & \bI_{c \times c}
\end{array} \right).
\ee
Observe that $\rs{\H_{\rm aug}}$ is purely isotropic.
The codespace is now described as the
simultaneous $+1$ eigenspace of
$$
\{  N_\bw: \bw \in \rs{\H_{\rm aug}}\},
$$
or, equivalently that of
$$
\cG_0 =  \{  N_\bw: \bw {\rm \,\,is \,\, a \,\, row \,\, of \,\,} \H_{\rm aug} \}.
$$
The decoding operation $\cD_0$ is also described in terms
of $\H$. The reduced syndrome
$\br = \H \odot \bu^\T$ is obtained by simultaneously measuring the observables in $\cG_0$.
The reduced error syndrome corresponds to a number of
possible errors $\bu \in S_0$ which all have an identical
effect on the codespace. Bob performs $\hat{N}_{\bu} = \hat{N}_{-\bu}$ to
undo the error.

%In fact, it suffices to resctrict the measurements
%to the set consisting of the \emph{rows} of $\H_I$ and $\H_S$
%as these generate the respective rowspaces.
%Observe that the parity check matrix $\H$
%merely serves as a compact description of $\rs{\H}$.
%It is the latter which completely determines everything.

%Note that $\ket{\bo}$ could have been replaced by $\ket{\bff}$.
%Then  parity check matrix  is the same. Different encoding.

\subsection{The general case}
\label{sec:general}
We now present our main result: how to convert an
arbitrary $(n + \hat{k})$-dimensional subspace $C$ of
$(\bbZ_2)^{2n}$ into a EAQEC code.
Consider the $(n - \hat{k})$-dimensional subspace $C^\perp$.
By Theorem \ref{thm1},
there exists a symplectic basis of $(\bbZ_2)^{2n}$
consisting of hyperbolic pairs $(\bu_i, \bv_i)$,
$i = 1, \dots ,n$, such that the ordered set
$\cR = \{\bu_{k + 1}, \dots, \bu_n, \bv_{k+ \ell +1}, \dots, \bv_{n}\}$
is a basis for $C^\perp$,
for some $c, \ell \geq 0$ with $2c + \ell = n - \hat{k}$,
and $k - c  = \hat{k}$.
Let $H$ be the matrix whose rows consist of the elements of $\cR$
in the order given from top to bottom.
Let $\Upsilon$ be the symplectomorphism defined by
\begin{eqnarray}
\Upsilon (\bu_i) &=& \bg_i \label{eq:gu} \\
\Upsilon (\bv_i) &=& \bh_i \label{eq:hv}.
\end{eqnarray}
Recall the matrix $\H$ given by
(\ref{uno1})-(\ref{uno3}).
Observe that, with a slight abuse of notation,
$$
\Upsilon(H) = \H
$$
in the sense that $\Upsilon$ takes the $i$th row of
$H$ to the $i$th row of $\H$.
We may extend $\Upsilon$ to act on $(\bbZ_2)^{2(n+c)}$,
including a trivial action on the bits corresponding to Bob's side.
Then
\be \label{lezaug}
\Upsilon(H_{\rm aug}) = \H_{\rm aug},
\ee
where $H_{\rm aug} = (H, B)$.

In terms of vector spaces
\begin{eqnarray}
\Upsilon(C^\perp) &=& C_0^\perp,  \label{cc01} \\
\Upsilon(\iso{C^\perp}) &=& \iso{C_0^\perp}. \label{cc02}
\end{eqnarray}
Note that
$c = \frac{1}{2}\dim \symp{C^\perp}$.
% and $n - \hat{k} =  \dim {C^\perp}$.
We our now ready for our main result:
\begin{theorem}
\label{thm3}
There exists an $[[n, k; c]]$ EAQEC code $(\hat{U}_{\rm enc}, \cD)$
with the following properties:
\begin{enumerate}
%\item It uses up $c$ ebits and encodes $k$ qubits.
%The net catalytic yield is
%$$
%K = n - \dim {C(H)^\perp}.
%$$
\item It can correct the
error set $S$ defined by: if $\bu, \bu' \in S$ and $\bu \neq \bu'$,
then either
\begin{enumerate}[i)]
\item $\bu - \bu' \not\in C$
(equivalently:
 $H \odot (\bu - \bu')^\T \neq \bo^\T$), or
\item $\bu - \bu' \in \iso{C^\perp}$
(equivalently: $\bu - \bu' \in \rs{H_I}$).
\end{enumerate}
\item  The codespace $\cC = \hat{U}_{\rm enc}(\cH^{\otimes k})$ is a
simultaneous  eigenspace of the ordered set
$$
\cG =  \{  N_\bw: \bw {\rm \,\,is \,\, a \,\, row \,\, of \,\,} H_{\rm aug} \},
$$
where $H_{\rm aug} = (H,B)$, with $B$ given by  (\ref{eq:beba}).
\item To decode, the reduced error syndrome
\be
\br = H \odot \bu^\T
\label{crnac}
\ee
is obtained by
simultaneously measuring the observables from $\cG$.
Bob finds a $\bu$ satisfying (\ref{crnac}) and
performs $\hat{N}_{\bu}$ to undo the error.
%In fact, it suffices to rectrict the measurements
%to the set consisting of the \emph{rows} of $\H_I$ and $\H_S$
%as these generate the respective rowspaces.
\end{enumerate}
\end{theorem}

\begin{figure}
\centerline{ {\scalebox{0.50}{\includegraphics{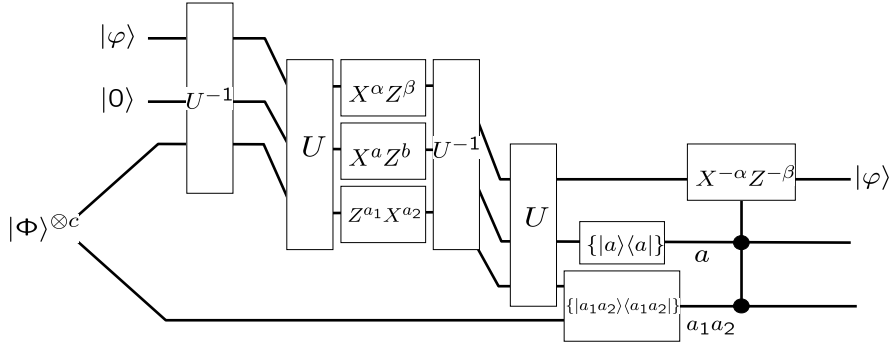}}}}
\caption{Generalizing the canonical code construction.}
\label{fignodumb}
\end{figure}

\noindent {\bf Remark}\,\, The above theorem generalizes
the error correcting conditions of \cite{Got98, CRSS98} for quantum
error correcting codes unassisted by entanglement.
When $c = 0$ then $C^\perp = \iso{C^\perp}$ and
no entanglement is used in the protocol. We call such codes
\emph{dual-containing}.

\medskip

\begin{proof}
%and the conditions from Proposition \ref{ecc} are
%invariant under symplectomorphisms.
By Theorem \ref{thm2} there exists a unitary
$U_\Upsilon$ such that for all $\bu \in (\bbZ_2)^{2n}$
\be
[N_{\Upsilon(\bu)}] = [U_\Upsilon N_{\bu} U_\Upsilon^{-1}],
\label{bulja}
\ee
and hence
$$
\hat{N}_{\Upsilon(\bu)} = \hat{U}_\Upsilon
\circ  \hat{N}_{\bu} \circ  \hat{U}_\Upsilon^{-1}.
$$
The above also holds for $\Upsilon$ and $ \hat{U}_\Upsilon$
extended to act trivially on Bob's side.

Our EAQEC code is defined by
$U_{\rm enc} = U_\Upsilon^{-1}  U_0$ and
$\cD = \cD_0 \circ \hat{U}_\Upsilon$,
%Alice performs the trivial encoding operation (\ref{tren})
%followed by $U_\Upsilon^{-1}$
 as shown in Figure \ref{fignodumb}.
\begin{enumerate}
\item Recall the error set $S_0$ defined in Proposition \ref{ecc}.
From (\ref{cc01}) and (\ref{cc02}) it follows that
$\Upsilon (S) =  S_0$.
By Proposition \ref{ecc}, for all $\bu \in S$,
$$
\cD_0 \circ \hat{N}_{\Upsilon (\bu)} \circ \hat{U}_{0}
 = \id^{\otimes k},
$$
from which
$$
\cD \circ \hat{N}_{\bu} \circ  \hat{U}_{\rm enc} = \id^{\otimes k}
$$
follows. Thus, the code $(\hat{U}_{\rm enc}, \cD)$ corrects the error set $S$.
\item The codespace is $\cC = U_\Upsilon^{-1} (\cC_0)$, by definition.
According to (\ref{lezaug}),
$\cC_0$ is the
simultaneous $+1$ eigenspace of
$$
\cG_0 = \{  N_{\Upsilon(\bw)}: \bw {\rm \,\,is \,\, a \,\,
row \,\, of \,\,} H_{\rm aug} \},
$$
or by (\ref{bulja}), the set
$$
\cG'_0 = \{   U_\Upsilon  N_{\bw}  U_\Upsilon^{-1}
  : \bw {\rm \,\,is \,\, a \,\,
row \,\, of \,\,} H_{\rm aug} \}.
$$
%$N_{{\rm row}(F,i)}$
Lemma \ref{simi} now implies that the codespace $\cC$ is a
simultaneous  eigenspace of $\cG$.
\item Assume that  error $\bu \in S$ occurs.
The operation $\cD_0$ involves
\begin{enumerate}
\item measuring the set of operators given by $\cG_0$, or equivalently $\cG'_0$,
yielding the reduced syndrome
$$
\br =  \H \odot  \Upsilon(\bu)^\T;
$$
\item performing $\hat{N}_{\Upsilon(\bu)}$,
where  $\Upsilon(\bu) \in S_0$ is an error
consistent with the observed syndrome $\br$.
\end{enumerate}
(\ref{crnac}) holds because
$$
\br =  \Upsilon(H) \odot  \Upsilon(\bu)^\T =
H \odot \bu^\T.
$$
By lemma \ref{simi2}, performing
$\cD = \cD_0 \circ \hat{U}_\Upsilon$ is
equivalent to
measuring the set $\cG$, followed by performing
$\hat{N}_\bu = \hat{U}^{-1}_\Upsilon
\circ  \hat{N}_{\Upsilon(\bu)} \circ  \hat{U}_\Upsilon$,
followed by $\hat{U}_\Upsilon$ to undo the encoding.
If the final $\hat{U}_\Upsilon$ is omitted, one recovers the
encoded state rather than the original one.

%As the pre-measurement state is the same as that in
%By the results of the previous subsection,
%for all $\bu \in S$,

\end{enumerate}
\end{proof}

\begin{lemma} \label{simi}
If $\cC_0$  is a
simultaneous  eigenspace of
Pauli operators from the set $\cG'_0$
then $\cC = U^{-1} (\cC_0)$
is a simultaneous  eigenspace of
Pauli operators  from the set
$\cG = \{ U^{-1} \bA U : \bA \in \cG'_0 \}$.
\end{lemma}

\begin{proof}
Observe that if
$$
\bA \ket{\psi} = \alpha  \ket{\psi},
$$
then
$$
(U^{-1} \bA U) U^{-1} \ket{\psi}
= \alpha  U^{-1} \ket{\psi}.
$$
\end{proof}

\begin{lemma} \label{simi2}
Performing $U$ followed by  measuring the operator
${\bf A}$ is equivalent to measuring the
operator $ U^{-1}{\bf A} U$ followed by performing $U$.
\end{lemma}

\begin{proof}
Let $\Pi_i$ be a projector onto the eigenspace corresponding
to eigenvalue $\lambda_i$ of ${\bf A}$.
Performing $U$ followed by  measuring the operator
${\bf A}$ is equivalent to the instrument (generalized measurement) given
by the set of operators $\{ \Pi_i U \}$.
The operator $ U^{-1}{\bf A} U$ has the same eigenvalues as $\bf A$,
and the projector onto the eigenspace corresponding
to eigenvalue $\lambda_i$ is $U^{-1} \Pi_i U$.
Measuring the operator $ U^{-1}{\bf A} U$ followed by performing $U$
is equivalent to the instrument
$\{U (U^{-1} \Pi_i U) \} = \{ \Pi_i U \}$.
\end{proof}

\subsection{Distance}
The notion of {distance} provides a convenient way to
characterize the error correcting properties of a code.
We start by defining the \emph{weight} of a vector $\bu  = (\bz| \bx)
\in (\bbZ_2)^{2n}$ by $\wt(\bu) = \wt(\bz \vee \bx)$.
Here $\vee$ denotes the bitwise logical ``or'',
and $\wt(\by)$ is the number
of non-zero bits in $\by \in (\bbZ_2)^{n}$.
% defined by $i \vee j = 1$ except when $i = j =
In terms of the Pauli group, $\wt(\bu)$ is the number of
single qubit Pauli matrices in $N_\bu$ not equal to the identity $I$.

Consider a symplectic code $C$.
% with parity check matrix $H$, so that $C^\perp = \rs{H}$.
The \emph{distance} of  $C$ is the maximum $d$ such that
for each nonzero $\bu$ of weight $< d$ either
\begin{enumerate}[i)]
\item $\bu  \not\in C$, or
%(equivalently:  $H \odot \bu^\T \neq \bo^\T$)
\item $\bu \in \iso{C^\perp}$
%(equivalently: $\bu  \in \rs{H_I}$)
\end{enumerate}

It is called \emph{non-degenerate} if the second condition
is not invoked.
A code is said to correct $t$ errors if it corrects the error
set $ \{ \bu: \wt(\bu) \leq t \}$ but not
 $ \{ \bu: \wt(\bu) \leq t+1 \}$.
Comparing these definitions with Theorem \ref{thm3}, a code
with distance $d = 2t + 1$ can correct $t$ errors.
% iff it has distance $d$.
An  $[[n,k;c]]$ EAQEC code with distance $d$
will be referred to as an $[[n,k,d;c]]$ code.

\section{Relation to quaternary codes}
\label{sec:quat}
We shall now show how to construct non-degenerate EAQEC
codes from classical codes over $\bbF_4$, generalizing the work
of \cite{CRSS98}. Following the presentation of Forney et al. \cite{FGG05},
the addition table of the additive group of the quaternary field
$\bbF_4 = \{0, 1, \omega, \overline{\omega}\}$ is given by
$$
\begin{array}{|c|cccc|}  \hline
+  & 0 & \overline{\omega} & 1 & \omega \\  \hline
0 & 0 &  \overline{\omega} & 1 & {\omega} \\
\overline{\omega} & \overline{\omega} & 0 & \omega & 1\\
1 & 1 & {\omega} & 0 &  \overline{\omega} \\
{\omega}& \omega & 1 & \overline{\omega} & 0   \\ \hline
\end{array}
$$
Comparing the above to  the addition table of $(\bbZ_2)^2$
establishes
the isomorphism $\gamma: \bbF_4 \rightarrow  (\bbZ_2)^2$,
given by the table
$$
\begin{array}{|c|c|} \hline
\bbF_4 & (\bbZ_2)^2 \\  \hline
 0 & 00\\
 \overline{\omega}  & 01\\
 1 & 11\\
 \omega  & 10\\ \hline
\end{array}
$$
The multiplication table for $\bbF_4$ is defined as
$$
\begin{array}{|c|cccc|}  \hline
\times  & 0 & \overline{\omega} & 1 & \omega  \\  \hline
0 & 0 & 0 & 0 & 0 \\
\overline{\omega} & 0 & \omega & \overline{\omega} & 1 \\
1 & 0 & \overline{\omega} & 1 & \omega \\
\omega & 0  & 1 & \omega & \overline{\omega} \\ \hline
\end{array}
$$
%In short, $0 + a = a$, $a + a = 0$ (so subtraction is the same as
%addition), and $1 + \omega + \overline{\omega} = 0$.
Define the \emph{traces} ($\tr$) of the elements
$\{0,
1, \omega, \overline{\omega}\}$ of $\bbF_4$ as
$\{0, 0, 1, 1\}$, and their \emph{conjugates} (``$^\dagger$'') as $\{0, 1,
\overline{\omega},  \omega\}$. Intuitively, $\tr a$ measures the
``$\omega$-ness'' of $a \in \bbF_4$.
Observe that $a  = 0$  if and only
if both  $\tr \omega a = 0$ and $\tr  \overline{\omega} a = 0$.
The \emph{Hermitian inner product} of two elements $a, b \in \bbF_4$ is
defined as $\inner{a}{b} = a^\dagger b \in \bbF_4$.
The \emph{trace product} is defined as
$\tr \inner{a}{b} \in \bbF_2$.  The trace  product table
is readily found to be
$$
\begin{array}{|c|cccc|}  \hline
\tr\inner{\,}{}  & 0 & \overline{\omega}& 1 & \omega  \\  \hline
0 & 0 & 0 & 0 & 0 \\
\overline{\omega} & 0& 0 & 1 & 1  \\
1 & 0& 1 & 0 & 1  \\
\omega & 0& 1 & 1 & 0  \\
\hline
\end{array}
$$
Comparing the above to  the $\odot$ table of $(\bbZ_2)^2$
establishes the identity
$$
\tr\inner{a}{b} = {\gamma(a) \odot \gamma(b)}.
$$
These notions can be generalized to $n$-dimensional vector spaces over
$\bbF_4$. Thus, for $\ba, \bb \in (\bbF_4)^n$,
\be
\tr\inner{\ba}{\bb} = {\gamma(\ba) \odot \gamma(\bb)^\T}.
\label{unx}
\ee
Let $\wt_4(\ba)$ be the number of non-zero bits in $\ba \in (\bbF_4)^{n}$.
Then we have another identity
\be
\wt(\gamma(\ba)) = \wt_4(\ba),
\label{deux}
\ee
where $\gamma(\ba) \in (\bbZ_2)^{2n}$.

\begin{proposition}
If a classical $[n,k,d]_4$ code exists then an
$[[n, 2k-n+c, d; c]]$ EAQEC code exists for some non-negative integer $c$.
\end{proposition}

\begin{proof}
Consider a classical $[n,k,d]_4$ code
(the subscript $4$ emphasizes that the code is over $\bbF_4$)
with an $(n - k) \times n$ quaternary parity check matrix $H_4$.
By definition, for each nonzero $\ba \in (\bbF_4)^n$ such
that $\wt_4(\ba) < d$,
$$
\inner{H_4}{\ba} \neq \bo^\T.
$$
This  is equivalent to the logical statement
$$
\tr \inner{\omega H_4}{\ba} \neq \bo^\T %{\rm \,\, or \,\,}
\,\,\vee\,\,
\tr \inner{\bar{\omega} H_4}{\ba} \neq \bo^\T.
$$
This is further equivalent to
$$
\tr \inner{\tilde{H}_4}{\ba} \neq \bo^\T,
$$
where
\be
\label{unos}
\tilde{H}_4 = \left(\begin{array}{c}
 \omega H_4 \\
\bar{\omega} H_4
\end{array} \right).
\ee
Define the $(2n - 2k) \times 2n$  symplectic matrix
$H = \gamma(\tilde{H}_4)$. By the correspondences
(\ref{unx}) and (\ref{deux}),
$$
H \odot \bu^\T \neq \bo^\T,
$$
holds for each nonzero $\bu \in (\bbZ_2)^{2n}$ with
$\wt(\bu) < d$.
Thus $C = \rs{H}^\perp$ defines a
non-degenerate $[n,2k-n+c,d;c]$ EAQEC code, where
$$
c = \frac{1}{2} \dim \symp{C}.
$$
%is the dimension of the purely
%symplectic subspace of $C$.
\end{proof}

Any classical binary $[n,k,d]_2$ code may be viewed as an
quaternary $[n,k,d]_4$. In this case, the above construction
gives rise to a CSS-type code.

\section{Catalytic quantum error correcting codes}

So far we have been considering \emph{communication} scenarios
involving two spatially separated parties Alice and Bob connected
by a noisy channel $\cN$. In this
setting, entanglement between them is a meaningful resource.
What if Alice and Bob are separated only in time?
For example, $\cN$ could represent the time evolution
of the state of a quantum computer.
Then  entanglement between Alice and Bob does not make sense any more.

\begin{figure}
\centerline{ {\scalebox{0.70}{\includegraphics{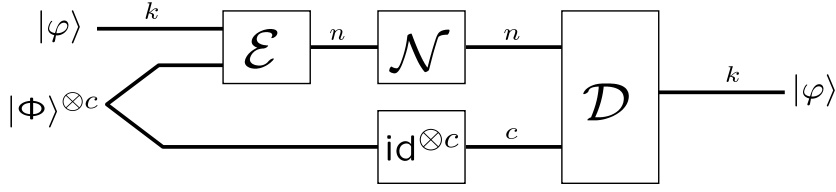}}}}
\caption{A catalytic quantum error-correcting code.}
\label{figcqecc}
\end{figure}

An alternative (in the communication picture) is to let Alice and Bob have access to a noiseless
channel, rather than entanglement. This channel serves as a catalyst
and is returned at the end of the protocol.
An $[[n,k;c]]$ \emph{catalytic} quantum error correcting (CQEC) code is
defined  by
\begin{itemize}
\item An encoding isometry $\cE: \cL^{\otimes k} \otimes \cL^{\otimes c}
 \rightarrow \cL^{\otimes n}$
\item A decoding CPTP map
$\cD: \cL^{\otimes n} \otimes \cL^{\otimes c} \rightarrow \cL^{\otimes k}$
\end{itemize}
such that
\be
 \cD   \circ (\cN \otimes \id^{\otimes c}) \circ \cE =
 \id^{\otimes {k}} = \id^{\otimes k-c} \otimes \id^{\otimes c}.
\label{ccon}
\ee
The above may be written as a resource inequality
\be
 \< \cN \>  + c \, [q \rightarrow q]  \geq
 (k-c) \, [q \rightarrow q]  + c\, [q \rightarrow q].
\ee
The net \emph{rate} of the CQEC code is given by $(k-c)/n$. 
Figure \ref{figcqecc} shows how any $[[n,{k};c]]$
 EAQEC code $(\cE, \cD)$ gives rise to a CQEC code with the same
parameters. This construction may be understood in terms of
resource inequalities. The simple protocol called
\emph{entanglement distribution} written as
$$
 c\, [q \rightarrow q] \geq  c\,[q \, q],
$$
creates $c$ ebits of entanglement by sending
half of a locally prepared
state $\ket{\Phi}^{\otimes c}$ through the channel $\id^{\otimes c}$.
The CQEC code is obtained by combining entanglement distribution with
the EAQEC code:
\be
\begin{split}
& \< \cN \>  + c \, [q \rightarrow q] \\
&\geq  \< \cN \>  + c \, [q \, q] \\
& \geq  k \, [q \rightarrow q] \\
&= (k-c) \, [q \rightarrow q]  + c\, [q \rightarrow q].
\end{split}
\ee
%with $\hat{k} = k - c$.
%The resource simulated by the protocol is the channel
%$$
%\id^{\otimes k} = \id^{\otimes c} \otimes \id^{\otimes \hat{k}}
%$$
%with $\hat{k} = k - c$.
%Thus $c$ qubit channels serve as a catalyst which is returned
%at the end of the protocol.
%This is the setting of \emph{catalytic} QEC codes (CQEC codes).
%We have seen that any EAQEC code $(\cE, \cD)$ gives rise to a CQEC code.
%This is explicitly shown in Figure \ref{figcqecc}.

Assume now that Alice and Bob have access to $m$ copies of
the chanel $\cN$. Performing the CQEC protocol $m$ times
in parallel (i.e. using the code $(\cE^{\otimes m}, \cD^{\otimes m})$)
gives
$$
m \< \cN \>  + m  c \, [q \rightarrow q] \geq
m (k-c) \, [q \rightarrow q] + m c \, [q \rightarrow q].
$$
The size of the catalyst can actually be reduced from $m c$ to $c$:
\be
m \< \cN \>  + c \, [q \rightarrow q] \geq
m (k-c) \, [q \rightarrow q] + c \, [q \rightarrow q].
\label{reduk}
\ee
The proof is by induction. The statement is trivial for $m = 1$.
For the inductive step, assume true for $m$. Then
(\ref{reduk}) holds for $m + 1$:
\be
\begin{split}
& (m + 1) \< \cN \>  + c \, [q \rightarrow q] \\
& =  \< \cN \>  + m \< \cN \>  + c \, [q \rightarrow q] \\
& \geq  \< \cN \>  + m (k-c) \, [q \rightarrow q] + c \, [q \rightarrow q]\\
& \geq m (k-c) \, [q \rightarrow q] +
(k-c) \, [q \rightarrow q] + c \, [q \rightarrow q].
\end{split}
\ee
A more conventional formulation of this catalyst reduction is given in the lemma below.
\begin{lemma}
If (\ref{ccon}) is satisfied then for any non-negative  integer $m$
there exists a CQEC code  $(\cE_m, \cD_m)$
for the channel $\cN^{\otimes m}$
in the sense that
$$
 \cD_m   \circ (\cN^{\otimes m} \otimes \id^{\otimes c}) \circ \cE_m =
 \id^{\otimes m (k-c)} \otimes \id^{\otimes c}.
$$
\end{lemma}

\begin{figure}
\centerline{ {\scalebox{0.40}{\includegraphics{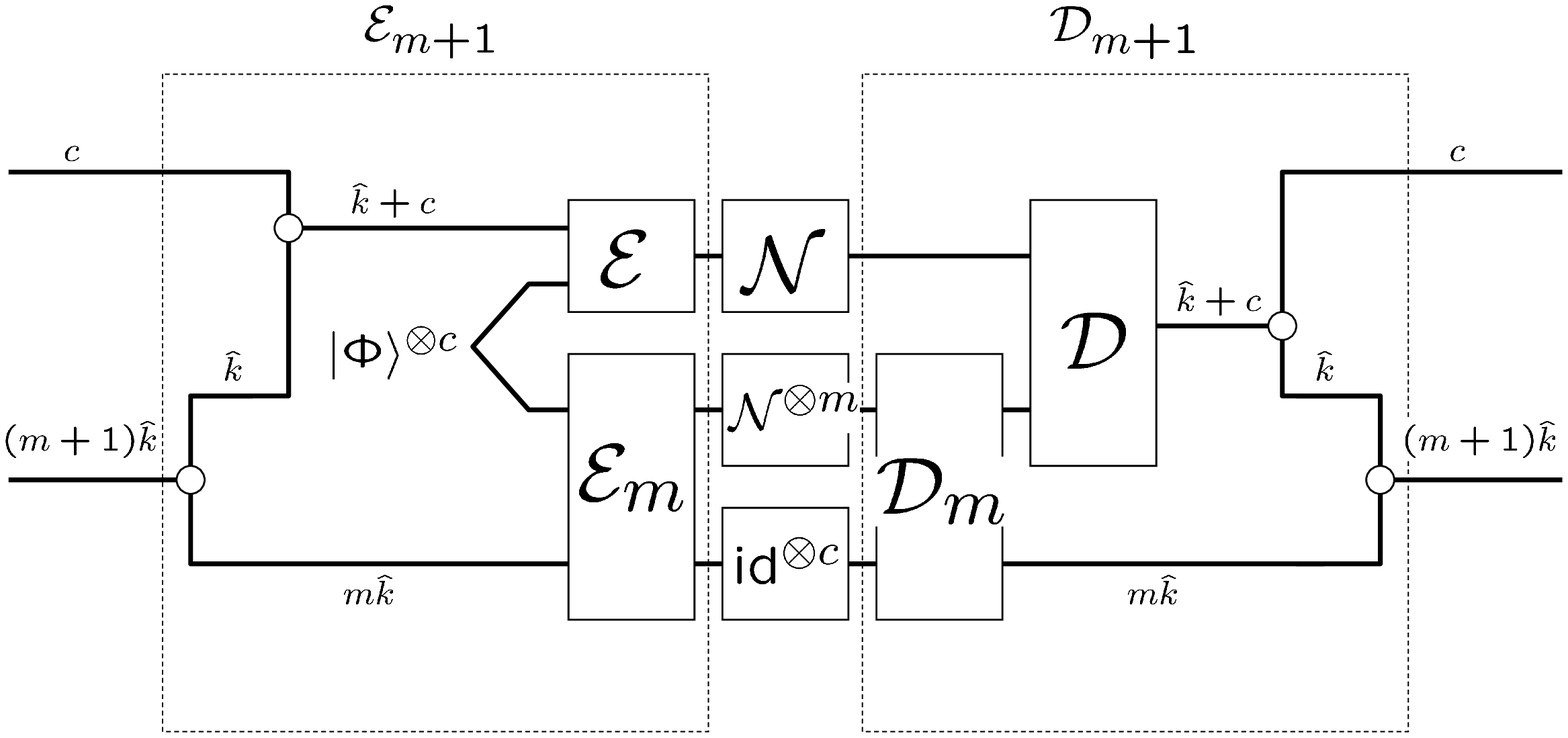}}}}
\caption{The inductive step.}
\label{figinduction}
\end{figure}

\begin{proof}
%By induction. The statement is trivial for $\ell = 0$.
The inductive step is shown in the Figure \ref{figinduction}.
\end{proof}

\medskip

The above construction is rather sensitive to perturbations.
If in any particular block a channel worse than $\cN$ is experienced,
the resulting channel will not be pure and the next block will
start with an impure catalyst.

\begin{figure}
\centerline{ {\scalebox{0.70}{\includegraphics{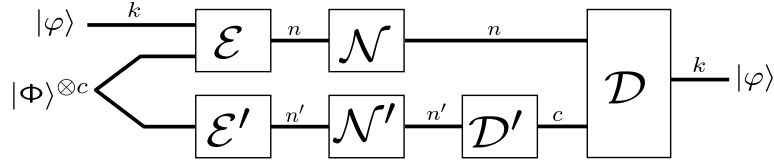}}}}
\caption{Constructing a QEC code from a seed QEC code and a CQEC code.}
\label{figseed}
\end{figure}

One may rightly ask about where one could obtain a catalyst to
begin with. %how does one ``borrow'' a perfect channel in practice.
A possibility is to use an ordinary $c=0$ QEC code. This is shown in Figure
\ref{figseed}. An $[n,{k};c]$ CQEC code for the channel $\cN$ combined
with a $[n',c; 0]$ QEC code for the channel $\cN'$ gives
an $[n + n',{k}; 0]$ QEC code for the channel $\cN \otimes \cN'$.
The combined code can be used as a catalyst for an even larger code.
In this way a sizeable catalyst can be built up pretty quickly.

It is worth looking at this construction from a purely
mathematical point of view. Let $C \subset (\bbZ_2)^{2n}$ and $C'\subset (\bbZ_2)^{2n'}$
be the symplectic codes corresponding to the
$[n,{k};c]$ CQEC code and $[n',c; 0]$ QEC code,
respectively. Let $H$ and $H'$ be the respective parity check matrices,
as in section \ref{sec:general}.
Note that ${C'}^\perp = \iso{{C'}^\perp}$.
Let  $\bu_i$, $i = 1, \dots, c$, be vectors in $(\bbZ_2)^{2n'}$ which,
together with a basis for ${C'}^\perp$, form a maximal $n'$-dimensional
isotropic subspace of $(\bbZ_2)^{2n'}$.
Recall the notation $\bg_i = (\bee_i| \bo) \in (\bbZ_2)^{2c}$. 
Let $\Upsilon$ be a symplectomorphism such that $\Upsilon(\bg_i) = \bu_i$.
%Let the hyperbolic pairs $(g_1,h_1), \dots, (g_c,h_c)$ form a basis for
%$C' \backslash {C'}^\perp$. Let $B_g$ be the $c \times 2n'$ matrix
%consisting of rows $g_1, \dots, g_c$, in that order.
%Let $B_h$ be the $c \times 2n'$ matrix
%consisting of rows $h_1, \dots, h_c$, in that order.
Define the $(n - {k}+c) \times 2n'$ matrix $B' = \Upsilon(B)$ with
$B$ defined as in (\ref{eq:beba}) and $\ell =  n - {k} - c$.
Note that the rows of $B'$ are in $C'$.
%\be B' =
% \left(\begin{array}{c}
% \bo_{m \times 2n'} \\
%  B_g \\
%  B_h
%\end{array} \right),
%\ee
%where $m = n - \hat{k} - 2c$.
Then
$$
\tilde{H}_{\rm aug} =
 \left(\begin{array}{cc}
H, & B' \\
\bo_{(n' - c) \times 2n}, & H'
\end{array} \right)
$$
is the parity check matrix for the combined
$[n + n',{k}; 0]$ QEC code. By construction, 
is must be dual-containing.

%Also: can combine different codes.
%$[n,k;0] + [n',k';k] + [n'',k''; k + k'] ...$

\section{A variation on EAQEC codes}

One lesson learned from quantum Shannon theory \cite{DHW03} is  that
catalytic and non-catalytic codes have similar performance.
In this section we mimic the quantum Shannon theoretical
construction from \cite{DHW03}.
First we construct codes for sending \emph{classical} information
with entanglement assistance. Then we make these protocols
\emph{coherent} in the sense of \cite{Har03, DHW03} to obtain
a variation on EAQEC codes in which entanglement is generated as well as
quantum communication. The end result is what we will call ``type II''
EAQEC codes, which can be constructed without the machinery of symplectic linear algebra.

\subsection{EA-codes for sending classical information}

The communication scenario again involves two spatially separated
parties, Alice and Bob.
The resources at their disposal are a noisy channel
$\cN: \cL^{\otimes n} \rightarrow \cL^{\otimes n}$ and
the shared $c$ ebit state $\ket{\Phi}^{\otimes c}$.
 Now Alice wishes to convey an element of $(\bbF_4)^k$  \emph{perfectly}
to Bob using the above resources.
A protocol which does this is called a $[n,k;c]_4$
entanglement-assisted classical error correcting code,
or EACEC code for short. We write the above as a resource inequality
\be
\label{EAcode}
\< \cN \> +  c \, [q \,q] \geq 2k \, [c \rightarrow c].
\ee
The factor of $2$ accounts for the conversion from
quaternary to binary.

Recall the isomorphism $\gamma: (\bbF_4)^n \rightarrow (\bbZ_2)^{2n}$.
It allows us to, with a slight abuse of notation,
speak of error sets  $S \subset (\bbF_4)^n$, and
Pauli matrices $N_\ba$, $\ba \in (\bbF_4)^n$.
Let $S \subset (\bbF_4)^{n}$ be the support of $\cN$.
An easy modification of Theorem \ref{discretization}
ensures that correctly decoding
the message for the set of channels
$\{\hat{N}_\ba: \ba \in S\}$ suffices for
the correct decoding of $\cN$.
The notion of distance for EACEC codes is equivalent to the
one for classical quaternary codes. An $[n,k;c]_4$ EACEC code of
distance $d$ is called a $[n,k,d;c]_4$ EACEC code.

%An $[n,k;n]_4$ entanglement-assisted classical error correcting code
%(EACECC) consists of
%\begin{itemize}
%\item An encoding map $(\bbF_4)^k \rightarrow \Pi^n$.
%Alice applies the image of this map to her half of
%$\ket{\Phi}^{\otimes n}$ which gets sent through the channel.
%\item A decoding measurement on $2n$ qubits
%with outcomes in $(\bbF_4)^k$.
%\end{itemize}

%The distance of the EACECC code is the maximum $d$ such that
%for each nonzero $\ba$ of quaternary weight $< d$,
%$H \odot \bu \neq \bo^\T$
%The code is called an $[n,k,d;n]_4$ code
%if it corrects the set

\begin{proposition}
If there exists an $[n,k]_4$ classical code (over $\bbF_4$) which corrects
the error set $S \subset (\bbF_4)^n$, then there exists an $[n,k;n]_4$ EACEC code
which corrects the same error set.
\end{proposition}

%\medskip

%\noindent{\bf Remark.} \, This relation also holds for nonlinear
%codes.

\begin{figure}
\centerline{ {\scalebox{0.60}{\includegraphics{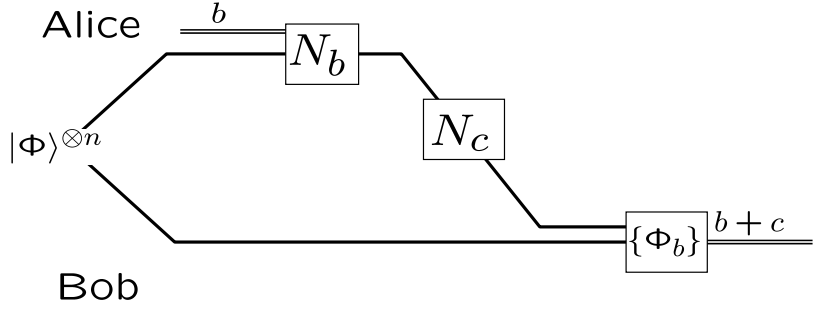}}}}
\caption{Reduction from an EACEC code to a classical code over $\bbF_4$.}
\label{figeaqecc}
\end{figure}

\begin{proof}
We will show that superdense coding
establishes an equivalence between a quantum Pauli error $N_{\bc}$
and a classical error $\bc$.

Assume $\bc = 0$, corresponding to no error. Alice
superdense encodes $\bb$ by performing $N_{\bb}$ on her half of
$\ket{\Phi}^{\otimes n}$. Bob performs a measurement
in the $\{ {\proj{\Phi_\bb}}: \bb \in (\bbF_4)^n \}$ basis,
where $\ket{\Phi_\bb} = (N_\bb \otimes I) \ket{\Phi}$,
thus decoding $\bb$.
%Let $H$ and $G$ be the parity check and generator
%matrices, respectively, of the $[n,k]_4$ classical code.
%Define $\bb = G \ba$.

If the channel is $\hat{N}_\bc$ for some $\bc \in S$, then
Alice's effective encoding becomes $N_\bc N_\bb$ which is
a representative of $[N_{\bb + \bc}]$.
Bob's measurement will reveal $\bb + \bc$ instead of $\bb$.
This is the message with a classical error $\bc \in S$.
The encoding preparation, followed by
quantum error $\hat{N}_\bc$ and decoding measurement, simulates
the noisy classical channel $\bb \mapsto \bb + \bc$.
The theorem now follows, since the classical code can correct any error $c \in S$ .
\end{proof}

\medskip

Thus there is a direct correspondence between
$[n,k,d]_4$ classical codes and $[n,k,d;n]$
EACEC codes. On the other hand, in
Section \ref{sec:quat} we saw  that an
$[n,k,d]_4$  classical code defines an
$[[n,2k - n+c,d;c]]$  EAQEC code.
In the next subsection we show how to
 construct a variation on an $[[n,2k - n+c,d;c]]$ EAQEC code
from an $[n,k,d;n]_4$ EACEC code via ``coherification.''

\subsection{Coherent EACEC codes}

At this point we need to introduce one more resource,
\emph{coherent communication} \cite{Har03}.
Let $\{\ket{0},\ket{1} \}$ denote a preferred basis for
a qubit system. The isometric channel which implements the
change of basis
$$
\Delta_2: \ket{i}^A \mapsto \ket{i}^A \ket{i}^B,  \,\, i = 0,1
$$
is called the \emph{coherent bit} (or \emph{cobit}) channel.
The superscript $A$ denotes a system held by Alice and
$B$ denotes a system held by Bob. It is regarded
as a coherent version of a classical bit channel.
Viewing it as a resource, we use the symbol
$[q \rightarrow  q \, q]$.
\emph{Coherifying} a protocol is a broad notion marked by
replacing classical communication by coherent communication
\cite{DHW03,Har03}. It was shown in \cite{Har03} that superdense
coding can be made coherent, i.e. that the following resource inequality
holds:
\be
[q \rightarrow q ]  + [q \, q] \geq 2 [q \rightarrow q \, q].
\ee

Consider an $[n,k,d;n]$ EACEC code, given by (\ref{EAcode}).
It can also be made coherent thanks to its connection to superdense
coding. In other words, (\ref{EAcode}) can be upgraded to
\be
\< \cN \>  +  n \, [q \,q] \geq 2k \, [q \rightarrow q \, q].
\label{eq:cohi}
\ee

\begin{figure}
\centerline{ {\scalebox{0.50}{\includegraphics{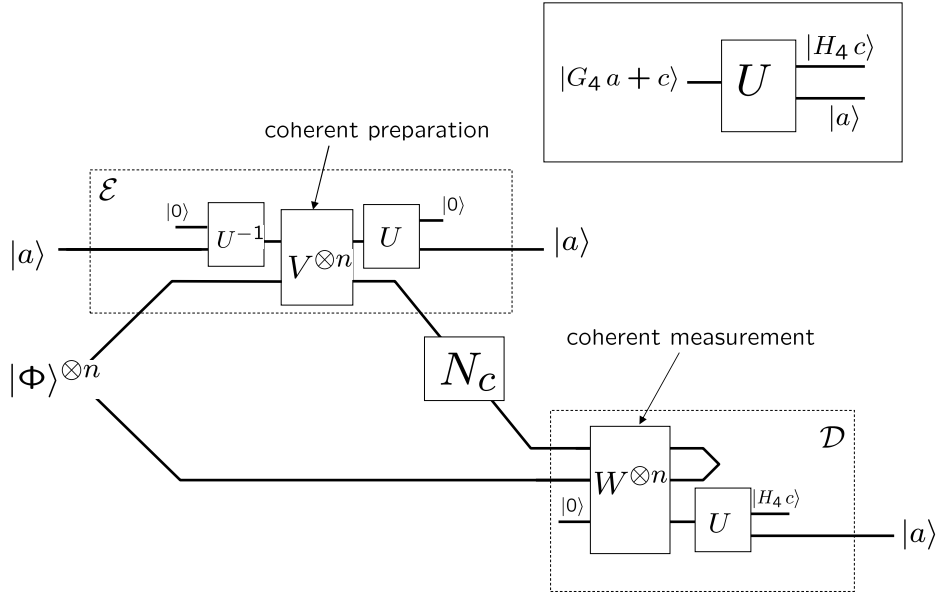}}}}
\caption{The circuit implementing a coherent EACEC code. The upper right hand corner
defines $U$ in terms of the quaternary code with parity check matrix $H_4$ and generating matrix $G_4$.}
\label{cohi}
\end{figure}

An explicit circuit implementing this resource inequality is
given in Figure \ref{cohi}.
The states $\{ \ket{\ba}:  \ba \in (\bbF_4)^k \}$ form
a basis for a $2k$ qubit space.
$\{ N_\bc \}$ is a  Pauli matrix whose index $\bc \in (\bbF_4)^n$
is in the support of $\cN$.
$H_4$ is the $(n-k) \times n$
quaternary parity check matrix for the classical $[n,k,d]_4$ code
which corrects all such $\bc$.
$G_4$ is the corresponding $n \times k$ generator matrix such
that $H_4 G_4 = {\bf 0}_{(n-k) \times k}$. The box in the upper right
hand corner defines the $4^n \times 4^n$ unitary matrix $U$.
%There $\bc \in (\bbF_4)^n$ represents a correctable error.
There $G_4 \ba \in (\bbF_4)^n$ is an encoded element $\ba$ of $(\bbF_4)^k$.
The unitaries $V$ and $W$ are given by
$V = \sum_{j \in \bbF_4} \ket{j}\bra{j}\otimes N_j$ and
$$W (\ket{\varphi} \ket{0})
= \sum_{j \in \bbF_4} \bra{\Phi_+}
(N_j^\dagger \otimes I) \ket{\varphi} \,
(\ket{\Phi_+} \ket{j}).$$

\begin{figure}
\centerline{ {\scalebox{0.50}{\includegraphics{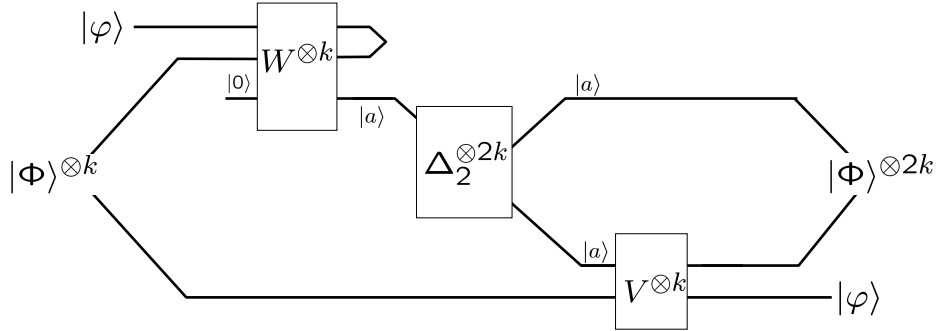}}}}
\caption{The circuit implementing coherent teleportation.}
\label{cotp}
\end{figure}

Harrow  \cite{Har03} also exhibited a coherent version of  quantum teleportation
\cite{BBCJPW93}, written as
\be
2k \, [q \rightarrow q \, q] + k \, [q \, q] \geq
k \, [q \rightarrow q ] + 2k  \,[q \, q].
\label{eq:cotp}
\ee
Figure \ref{cotp} depicts a circuit implementing this resource inequality.

Combining (\ref{eq:cohi}) with (\ref{eq:cotp}) gives
\be
\< \cN \>  +  (n +  k) \, [q \,q]
\geq k \, [q \rightarrow  q]
+ 2k  \,[q \, q].
\label{catdad}
\ee
This differs from a hypothetical $[[n, k; n - k]]$ EAQEC code\footnote{
This EAQEC code has the maximum value of $c = n - k$.}
given by
$$
\< \cN \>  +  (n -  k) \, [q \,q]
\geq k \, [q \rightarrow  q]
$$
in that an extra $2k \,[q \, q]$ is needed as a catalyst.
We call this a \emph{type II}  $[[n, k ; n - k; 2k]]$
EAQEC code, and will refer to the EAQEC codes from Section 2 as
\emph{type I} EAQEC codes. 
A  EAQEC code is not as versatile as regular EAQEC codes.
The catalyst does not allow it to be converted into a catalytic QEC code, for example.
Also, type II EAQEC codes appear to be limited to $\bbF_4$ construction.

As in the original Shannon theoretical result \cite{DHW03} (Figure \ref{famfig}),
 type II  EAQEC
codes (\ref{catdad}) can be combined with superdense coding (\ref{eq:sd})
to give a catalytic version of an EACEC code (\ref{EAcode}):
$$
\< \cN \>  +  n \, [q \,q]   + k  \, [q \,q]
\geq 2k \, [c \rightarrow  c] + k  \,[q \, q].
$$
This does not  hold for type I  EAQEC codes of
Section \ref{sec:EAQECcode}, unless $c$  equals its maximal value of $n-k$.

\section{Bounds on performance}

In this section we shall see that the performance of
EAQEC codes is comparable to the performance of
QEC codes (which are a special case of EAQEC codes).

The two most important outer bounds for QEC codes are the
quantum Singleton bound \cite{KL97, Pre98}
and the quantum Hamming bound \cite{Got96}.
Given an $[[n,k,d]]$ QEC code (which is
an $[[n,k,d;0]]$ EAQEC code), the
quantum Singleton bound reads
$$
n - k \geq 2 (d-1).
$$
The quantum Hamming  bound holds only for non-degenerate codes
and reads
$$
\sum_{j =0}^{\lfloor \frac{d-1}{2} \rfloor} 3^j {n \choose j}
\leq 2^{n - k}.
$$
The proofs of these bounds \cite{Pre98,Got96} are easily adapted to
EAQEC codes with $k-c$ playing the role of $k$.
This was first noted by Bowen \cite{Bow02} in the case of the
quantum Hamming bound, which now reads
$$
\sum_{j =0}^{\lfloor \frac{d-1}{2} \rfloor} 3^j {n \choose j}
\leq 2^{n - k+c}.
$$
The  new quantum  Singleton bound is 
$$
n - k + c \geq 2 (d-1).
$$
%Consequently, an $[[n,k,d;c]]$ EAQEC code
%satisfies both bounds for any value of $c$.
Note that the $\bbF_4$ construction connects the quantum
Singleton bound to the classical Singleton bound
$n-k \geq d-1$. An $[n,k,d]_4$ code
saturating the classical Singleton bound implies
an $[[n,2k - n + c, d; c]]$ EAQEC code saturating the
quantum Singleton bound.

It is instructive to examine the asymptotic performance
of quantum codes on a particular channel.
A popular choice is the tensor power channel
$\cN^{\otimes n}$, where $\cN$ is the depolarizing channel
with Kraus operators $\{ \sqrt{p_0} I,
\sqrt{p_1} X, \sqrt{p_2} Y, \sqrt{p_3} Z \}$,
for some probability vector
${\bf p} = (p_0, p_1,  p_2, p_3 )$.
%We say that a rate $R = k/n$ is achievable by
%a QEC code on the channel $\cN^{\otimes n}$
%if there exists an $[[n,k]]$ code that
%corrects all the typical errors.

It is well known that the maximal transmission
rate $R = k/n$ achievable by a non-degenerate QEC code
(in the sense of vanishing error for large $n$ on the
channel $\cN^{\otimes n}$) is equal to the
\emph{hashing bound} $R = 1 - H({\bf p})$.
Here $H({\bf p})$ is the Shannon entropy of the probability
distribution ${\bf p}$.
This bound is attained by picking a random dual-containing code.
However no explicit constructions are known which achieve this bound.

Interestingly, the $\bbF_4$ construction also connects the
hashing bound to the Shannon bound for quaternary channels.
Consider the quaternary channel $a \mapsto a + c$, where
$c$ takes on values $0, \omega,  1, \bar{\omega}$,
with respective probabilities $p_0, p_1,  p_2, p_3$.
The maximal achievable rate $R = k/n$ for this channel
was proved by Shannon to equal $R = 2 - H({\bf p})$.
An $[n,k]_4$ code saturating the Shannon bound implies
an $[[n,2k - n+c; c]]$ CQEC code achieving the hashing bound! 
(Recall that the net rate of an $[[n,k;c]]$ CQEC code is defined as 
$R = (k-c)/n$.)
Efficiently decodable modern classical codes, such as low-density parity check (LDPC) codes
and turbo codes, are known to come very close to the Shannon bound.
These codes are not at all guaranteed to be dual-containing \cite{MMM04}, which
means that no hashing bound attaining QEC code can be constructed from them.
However, a  \emph{catalytic} QEC code can. Directly investigating
the symplectic structure of such modern codes will reveal the size of the catalyst.
We expect that bootstrapping the method from Figure \ref{figseed} will
enable us to construct a QEC code with similar properties.

\section{The [[3,1,3;2]] EAQEC code}
In this section, we will construct a
$[[3,1,3;2]]$ EAQEC code and relate this code to Bowen's earlier result \cite{Bow02}.
Consider the classical $[3,1,3]$ quaternary code with parity check
matrix
\be H_4 = \left(\begin{array}{ccc}
1 & 1 & 0 \\
1 & 0 & 1
\end{array} \right).
\ee Then \be H=\gamma(\tilde{H}_4) = \left(\begin{array}{ccc|ccc}
1 & 1 & 0 & 0 & 0 & 0 \\
1 & 0 & 1 & 0 & 0 & 0 \\
0 & 0 & 0 & 1 & 1 & 0 \\
0 & 0 & 0 & 1 & 0 & 1
\end{array} \right).
\ee Following the proof of Theorem 1.1, we have \be \begin{array}{ccccccc}
\bu_1  = &( 1 & 1 & 0 & 0 & 0 & 0) \\
\bu_2  = &( 0 & 0 & 0 & 1 & 1 & 0) \\
\bu_3  = &( 1 & 1 & 1 & 0 & 0 & 0) \\
\bv_1  = &( 0 & 0 & 0 & 1 & 0 & 1) \\
\bv_2  = &( 1 & 0 & 1 & 0 & 0 & 0) \\
\bv_3  = &( 0 & 0 & 0 & 1 & 1 & 1),
\end{array}
\ee and the hyperbolic pairs $(\bu_1,\bv_1)$ and $(\bu_2,\bv_2)$ span the
rowspace of $H$.
%The symplectomorphism $\Upsilon$ is
%\be
%\Upsilon= \left(\begin{array}{cccccc}
%1 & 1 & 0 & 0 & 0 & 0 \\
%1 & 0 & 1 & 0 & 0 & 0 \\
%0 & 0 & 0 & 1 & 1 & 0 \\
%0 & 0 & 0 & 1 & 0 & 1
%\end{array} \right).
%\ee
The simultaneous $+1$ eigenstate of the commuting operators
$N_{\bu_i}$, $i=1,2,3$, is
$$\ket{\tilde{000}}=\frac{1}{\sqrt{2}}(\ket{000}+\ket{110}).$$ Then
\be
\begin{split}
\ket{\tilde{000}}&=\frac{1}{\sqrt{2}}(\ket{000}+\ket{110}) \\
\ket{\tilde{001}}&=N_{\bv_1}\ket{\tilde{000}}=(X\otimes I\otimes X)\ket{\tilde{000}}=\frac{1}{\sqrt{2}}(\ket{101}+\ket{011})\\
\ket{\tilde{010}}&=N_{\bv_2}\ket{\tilde{000}}=(Z\otimes I\otimes Z)\ket{\tilde{000}}=\frac{1}{\sqrt{2}}(\ket{101}-\ket{011})\\
\ket{\tilde{011}}&=N_{\bv_1+\bv_2}\ket{\tilde{000}}=(Y\otimes I\otimes Y)\ket{\tilde{000}}=\frac{1}{\sqrt{2}}(-\ket{101}+\ket{011})\\
\ket{\tilde{100}}&=N_{\bv_3}\ket{\tilde{000}}=(X\otimes X\otimes X)\ket{\tilde{000}}=\frac{1}{\sqrt{2}}(\ket{111}+\ket{001})\\
\ket{\tilde{101}}&=N_{\bv_1+\bv_3}\ket{\tilde{000}}=(I\otimes X\otimes I)\ket{\tilde{000}}=\frac{1}{\sqrt{2}}(\ket{010}+\ket{100})\\
\ket{\tilde{110}}&=N_{\bv_2+\bv_3}\ket{\tilde{000}}=(Y\otimes X\otimes Y)\ket{\tilde{000}}=\frac{1}{\sqrt{2}}(-\ket{111}+\ket{001})\\
\ket{\tilde{111}}&=N_{\bv_1+\bv_2+\bv_3}\ket{\tilde{000}}=(Z\otimes X\otimes
Z)\ket{\tilde{000}}=\frac{1}{\sqrt{2}}(\ket{010}-\ket{100})
\end{split}
\ee The encoding unitary $U_{\Upsilon}$ is therefore \be
U_{\Upsilon} = \frac{1}{\sqrt{2}} \left(\begin{array}{cccccccc}
1 & 0 & 1 & 0 & 0 & 0 & 0 & 0 \\
0 & 0 & 0 & 0 & 1 & 0 & 1 & 0 \\
0 & 0 & 0 & 0 & 0 & 1 & 0 & 1 \\
0 & 1 & 0 & 1 & 0 & 0 & 0 & 0 \\
0 & 0 & 0 & 0 & 0 & 1 & 0 & -1 \\
0 & 1 & 0 & -1 & 0 & 0 & 0 & 0 \\
1 & 0 & -1 & 0 & 0 & 0 & 0 & 0 \\
0 & 0 & 0 & 0 & 1 & 0 & -1 & 0
\end{array} \right).
\ee The logical $0$ and $1$  codewords are \be
\begin{split}
\ket{0_L}&=U_{\Upsilon} \ket{0}\ket{\Phi_+}^{\otimes 2}
=\frac{1}{2}\left(\ket{\tilde{000}}\ket{00}+\ket{\tilde{001}}\ket{01}+\ket{\tilde{010}}\ket{10}+\ket{\tilde{011}}\ket{11}\right) \\
\ket{1_L}&=U_{\Upsilon} \ket{1}\ket{\Phi_+}^{\otimes
2}=\frac{1}{2}\left(\ket{\tilde{100}}\ket{00}+\ket{\tilde{101}}\ket{01}+\ket{\tilde{110}}\ket{10}+\ket{\tilde{111}}\ket{11}\right)
\end{split}
\ee
Bowen's code \cite{Bow02} can be obtained by applying the following
unitary to the codewords given above \be U_{B} =
\frac{1}{2\sqrt{2}} \left(\begin{array}{cccccccc}
1 & -1 & -1 & -1 & -1 & 1 & 1 & 1 \\
-1 & -1 & -1 & 1 & 1 & 1 & 1 & -1 \\
1 & -1 & 1 & 1 & -1 & 1 & -1 & -1 \\
-1 & -1 & 1 & -1 & 1 & 1 & -1 & 1 \\
1 & -1 & -1 & -1 & 1 & -1 & -1 & -1 \\
1 & 1 & 1 & -1 & 1 & 1 & 1 & -1 \\
-1 & 1 & -1 & -1 & -1 & 1 & -1 & -1 \\
-1 & -1 & 1 & -1 & -1 & -1 & 1 & -1
\end{array} \right).
\ee
%The encoding unitary $U_B$ in Bowen's code is
%\be
%U_{B} = \frac{1}{2} \left(\begin{array}{cccccccc}
%1 & 0 & 0 & -1 & 0 & -1 & -1 & 0 \\
%0 & 1 & -1 & 0 & -1 & 0 & 0 & -1 \\
%0 & 1 & 1 & 0 & -1 & 0 & 0 & 1 \\
%-1 & 0 & 0 & -1 & 0 & 1 & -1 & 0 \\
%0 & -1 & 1 & 0 & -1 & 0 & 0 & -1 \\
%1 & 0 & 0 & -1 & 0 & 1 & 1 & 0 \\
%-1 & 0 & 0 & -1 & 0 & -1 & 1 & 0 \\
%0 & -1 & -1 & 0 & -1 & 0 & 0 & 1
%\end{array} \right).
%\ee

\section{Table of codes}

In \cite{CRSS98} a table of best known QEC codes was given.
Below we show an updated table which includes EAQEC codes.

\bigskip

\begin{tabular}{|c||c|c|c|c|c|c|c|c|c|c|c|}
  \hline
  $n \backslash {k-c}$ & 0           & 1            & 2            & 3            & 4             & 5           & 6           & 7    & 8   & 9    & 10      \\ \hline
  3              & $2$       & $\,\,\,2^*$     & $1$    & $1$     &               &             &             &       & & &     \\ \hline
  4              & $\,\,\,3^*$       & $2$    & $2$     & $1$    & $1$      &             &             &        & & &      \\ \hline
  5              & $3$   & $3$     & $2$     & $\,\,\,2^*$        & $1$     & $1$    &             &        & & &      \\ \hline
  6              & $4$       & $3$      & $2$     & $2$     & $2$         & $1$   & $1$    &         & & &     \\ \hline
  7              & $3$       & $3$     & $2$    & $2$     & $2$      & $\,\,\,2^*$       & $1$   & $1$   &  & &  \\ \hline
  8              & $4$       & $3$    & $3$     & $3$        & $2$      & $2$    & $2$       & $1$   &   $1$ &  &  \\ \hline
  9              & $4$   & $\,\,\,4^*$     & $3$    & $3$     & $2$     & $2$    & $2$    & $\,\,\,2^*$     &  $1$   &  $1$ &   \\ \hline
  10             & $\,\,\,5^*$       & $4$    & $4$     & $3$    & $3$      & $2$   & $2$    & $2$   & $2$   & $1$   &  $1$  \\ \hline

  \hline
\end{tabular}

\bigskip

The entries with an asterisk mark the improvements over the table from \cite{CRSS98}. 
All these are obtained from  Proposition 3.1. The corresponding classical 
quaternary code is available online at
{\tt http://www.win.tue.nl/$\sim$aeb/voorlincod.html}.

The general methods from \cite{CRSS98} for constructing new codes
from old also apply here. Moreover, new constructions are possible
since the dual-containing condition is lifted.
An example is  given by the following Theorem.

\begin{theorem}
 a) Suppose an $[[n,{k},d;c]]$ code exists, then
an $[[n+1,{k}-c + c'-1,d';c']]$ code exists for some $c'$ and $d' \geq d$; 
b) Suppose a non-degenerate $[[n,{k},d;c]]$ code exists, then
an $[[n-1,{k}-c + c' +1,d-1;c']]$ code exists for some $c'$.
\end{theorem}

\begin{proof}
a) Let $H$ be the $(n-{k} + c \times 2n)$ parity check
matrix of the $[[n,{k},d;c]]$ code. The parity check matrix of 
the new $[[n+1,{k}-c + c'-1,d';c']]$ is then
\be H'=\left(\begin{tabular}{cc|cc}
  0 $\cdots$ 0 & 0 & 1 $\cdots$ 1 & 1\\
  1 $\cdots$ 1 & 1 & 0 $\cdots$ 0 & 0\\
   & 0  &  &0\\
   \raisebox{0.5ex}[0cm][0cm]{$H_Z$} & \vdots & \raisebox{0.5ex}[0cm][0cm]{$H_X$} & \vdots \\
   & 0  &  &0
\end{tabular} \right).
\ee
This corresponds to the classical construction of adding a parity check at the end of the codeword  \cite{FJM77}.
The additional rows ensure that errors involving the last qubit are detected. Sometimes the distance
actually increases: for instance, the
 $[[8,0,4;0]]$ code is obtained from the $[[7,1,3;0]]$ code in this way.

b) We mimic the classical ``puncturing'' method \cite{FJM77}. 
Let $C$ be the $(n+{k}-c)$-dimensional subspace of $(\bbZ_2)^{2n}$ corresponding to the  
$[[n,{k},d;c]]$ 
EAQEC code. Puncturing $C$ by deleting the first $Z$ and $X$ coordinate, we obtain a new ``code'' $C'$
which is an $(n+{k}-c)$-dimensional subspace of  $(\bbZ_2)^{2(n-1)}$. This corresponds to an
$[[n-1,{k}-c+c'+1,d-1;c']]$ EAQEC code, as the minimum distance between the ``codewords'' of $C$ decreases
by at most $1$.
\end{proof}

\section{Discussion}

Motivated by recent developments in quantum Shannon theory,
we have introduced a generalization of the stabilizer formalism to
the setting in which the encoder Alice and decoder Bob pre-share entanglement
(EAQEC codes). We have traced the male side of family tree of quantum Shannon theory,
from EAQEC codes (corresponding to the father protocol) to catalytic quantum codes (corresponding to
the quantum capacity) and EACEC codes (corresponding to the classical EA-capacity). Moreover,
EACEC codes can be made coherent, providing an alternative to the EAQEC construction from Section 2.
The most obvious question is whether we can do the same for the female side of the
family tree \cite{DHW03}. Preliminary results \cite{LD06a} give a positive answer to this question:
entanglement distillation protocols assisted by quantum and classical communication can be constructed based 
on non-orthogonal symplectic codes.

There are two practical advantages of EAQEC codes over standard QEC codes:
\begin{enumerate}
\item They are much easier to construct from classical
codes because they are not required to be dual-containing.
This allows us to import the classical theory of error correction wholesale,
including capacity-achieving modern codes.
We plan to examine the performance of classical LDPC codes and turbo codes
in the context of the catalyst size for EAQEC codes.

\item The entanglement used in the protocol is a strictly weaker resource
than quantum communication. Thus comparing $[[n,k,d;c]]$ EAQEC codes
to $[[n,k-c,d;0]]$ QEC codes is not being entirely fair to former.
The pre-shared entanglement could have been obtained from a two way entanglement
distillation protocol which achieve higher rates compared than  one-way schemes.
In this sense, a large value of the catalyst $c$ is viewed as advantageous,
as it implies a higher qubit channel yield.
\end{enumerate}

If one is interested in applications to fault tolerant quantum computation,
where the resource of entanglement is meaningless, high values of $c$
are unwelcome because they require a long seed QEC code. We expect this obstacle
to be overcome by bootstrapping.

Another fruitful line of investigation connects to quantum cryptography.
Quantum cryptographic protocols, such as BB84, are intimately related
to CSS QEC codes. In \cite{LD06b} it is shown that EAQEC analogues
of CSS codes give rise to key expansion protocols which do not rely on the
existence of long dual-containing codes.

We thank Graeme Smith for pointing us to references \cite{BFG05} and \cite{FCY+04}.
 TAB acknowledges financial support from NSF Grant No.~CCF-0448658, and TAB and MHH both received support from NSF Grant No.~ECS-0507270.  ID and MHH acknowledge financial support from NSF Grant No.~CCF-0524811 and NSF Grant No.~CCF-0545845.

\bibliography{ref2}

\begin{thebibliography}{10}

\bibitem{BBCJPW93}
C.~H. Bennett, G.~Brassard, C.~Cr\'{e}peau, R.~Jozsa, A.~Peres, and W.~K.
  Wootters.
\newblock Teleporting an unknown quantum state via dual classical and
  {E}instein-{P}odolsky-{R}osen channels.
\newblock {\em Phys. Rev. Lett.}, 70, 1993.

\bibitem{BDSW96}
C.~H. Bennett, D.~P. DiVincenzo, J.~A. Smolin, and W.~K. Wooters.
\newblock Mixed state entanglement and quantum error correction.
\newblock {\em Phys. Rev. A}, 52:3824--3851, 1996.
\newblock quant-ph/9604024.

\bibitem{BSST01}
C.~H. Bennett, P.~W. Shor, J.~A. Smolin, and A.~Thapliyal.
\newblock Entanglement-assisted capacity of a quantum channel and the reverse
  {S}hannon theorem.
\newblock {\em IEEE Trans. Inf. Theory}, 48, 2002.
\newblock quant-ph/0106052.

\bibitem{BW92}
C.~H. Bennett and S.~J. Wiesner.
\newblock Communication via one- and two-particle operators on
  {E}instein-{P}odolsky-{R}osen states.
\newblock {\em Phys. Rev. Lett.}, 69:2881--2884, 1992.

\bibitem{Bow02}
G.~Bowen.
\newblock Entanglement required in achieving entanglement-assisted channel
  capacities.
\newblock {\em Phys. Rev. A}, 66:052313, 2002.
\newblock quant-ph/0205117.

\bibitem{BFG05}
S.~Bravyi, D.~Fattal, and D.~Gottesman.
\newblock {GHZ} extraction yield for multipartite stabilizer states.
\newblock {\em J. Math. Phys.}, 47:062106, 2006.
\newblock quant-ph/0504208.

\bibitem{CRSS97}
A.~R. Calderbank, E.~M. Rains, P.~W. Shor, and N.~J.~A. Sloane.
\newblock Quantum error correction and orthogonal geometry.
\newblock {\em Phys. Rev. Lett.}, 78:405--408, 1997.
\newblock quant-ph/9605005.

\bibitem{CRSS98}
A.~R. Calderbank, E.~M. Rains, P.~W. Shor, and N.~J.~A. Sloane.
\newblock Quantum error correction via codes over {G}{F}(4).
\newblock {\em IEEE Trans. Inf. Theory}, 44:1369--1387, 1998.
\newblock quant-ph/9608006.

\bibitem{CS96}
A.~R. Calderbank and P.~W. Shor.
\newblock Good quantum error-correcting codes exist.
\newblock {\em Phys. Rev. A}, 54:1098--1105, 1996.
\newblock quant-ph/9512032.

\bibitem{Silva01}
A.~C. da~Silva.
\newblock {\em Lectures on symplectic geometry}.
\newblock Springer-Verlag, Berlin, 2001.

\bibitem{Devetak03}
I.~Devetak.
\newblock The private classical capacity and quantum capacity of a quantum
  channel.
\newblock {\em IEEE Trans. Inf. Theory}, 51(1):44--55, 2005.
\newblock quant-ph/0304127.

\bibitem{DHW05}
I.~Devetak, A.~W. Harrow, and A.~Winter.
\newblock A resource framework for quantum {S}hannon theory, 2005.
\newblock quant-ph/0512015.

\bibitem{DHW03}
I.~Devetak, A.~W. Harrow, and A.~J. Winter.
\newblock A family of quantum protocols.
\newblock {\em Phys. Rev. Lett.}, 93, 2004.
\newblock quant-ph/0308044.

\bibitem{DW03a}
I.~Devetak and A.~Winter.
\newblock Distilling common randomness from bipartite quantum states.
\newblock {\em IEEE Trans. Inf. Theory}, 50:3138--3151, 2003.
\newblock quant-ph/0304196.

\bibitem{FCY+04}
D.~Fattal, T.~S. Cubitt, Y.~Yamamoto, S.~Bravyi, and I.~L. Chuang.
\newblock Entanglement in the stablizer formalism, 2004.
\newblock quant-ph/0406168.

\bibitem{Got96}
D.~Gottesman.
\newblock A class of quantum error correcting codes saturating the quantum
  {H}amming bound.
\newblock {\em Phys. Rev. A}, 54:1862, 1996.

\bibitem{Got98}
D.~Gottesman.
\newblock A theory of fault-tolerant quantum computation.
\newblock {\em Phys. Rev. A}, 57:127--137, 1998.
\newblock quant-ph/9702029.

\bibitem{Hamada02}
M.~Hamada.
\newblock Information rates achievable with algebraic codes on quantum discrete
  memoryless channels, 2002.
\newblock quant-ph/0207113.

\bibitem{Har03}
A.~W. Harrow.
\newblock Coherent communication of classical messages.
\newblock {\em Phys. Rev. Lett.}, 92, 2004.
\newblock quant-ph/0307091.

\bibitem{FGG05}
G.~D.~Forney Jr., M.~Grassl, and S.~Guha.
\newblock Convolutional and tail-biting quantum error-correcting codes, 2005.
\newblock quant-ph/0511016.

\bibitem{KL97}
E.~Knill and R.~Laflamme.
\newblock A theory of quantum error correcting codes.
\newblock {\em Phys. Rev. A}, 55:900, 1997.
\newblock quant-ph/9604034.

\bibitem{LMPZ96}
R.~Laflamme, C.~Miquel, J.-P. Paz, and W.~H. Zurek.
\newblock Perfect quantum error-correction code.
\newblock {\em Phys. Rev. Lett.}, 77:198--201, 1996.
\newblock quant-ph/9602019.

\bibitem{Lloyd96}
S.~Lloyd.
\newblock Capacity of the noisy quantum channel.
\newblock {\em Phys. Rev. A}, 55, 1996.
\newblock quant-ph/9604015.

\bibitem{LD06a}
Z.~Luo and I.~Devetak.
\newblock Catalytic entanglement distillation, 2006.
\newblock  In preparation.

\bibitem{LD06b}
Z.~Luo and I.~Devetak.
\newblock Quantum key expansion from non-orthogonal codes, 2006.
\newblock quant-ph/0608029.

\bibitem{MMM04}
D.~J.~C. MacKay, G.~Mitchison, and P.~L. McFadden.
\newblock Sparse-graph codes for quantum error correction.
\newblock {\em IEEE Trans. Inf. Theory}, 50:2315--2330, 2004.

\bibitem{FJM77}
F.J. MacWilliams and N.J.A. Sloane.
\newblock {\em The Theory of Error-Correcting Codes}.
\newblock Elsevier, Amsterdam, 1977.

\bibitem{NC00}
M.~A. Nielsen and I.~L. Chuang.
\newblock {\em Quantum Computation and Quantum Information}.
\newblock Cambridge University Press, New York, 2000.

\bibitem{Pre98}
J.~Preskill.
\newblock Lecture notes for physics 229: Quantum information and computation,
  1998.
\newblock http://www.theory.caltech.edu/people/preskill/ph229.

\bibitem{Shannon48}
C.~E. Shannon.
\newblock A mathematical theory of communication.
\newblock {\em Bell System Tech.~Jnl.}, 27:379--423, 623--656, 1948.

\bibitem{Sho95}
P.~W. Shor.
\newblock Scheme for reducing decoherence in quantum computer memory.
\newblock {\em Phys. Rev. A}, 52:2493--2496, 1995.

\bibitem{Shor02}
P.~W. Shor.
\newblock The quantum channel capacity and coherent information.
\newblock MSRI workshop on quantum computation, 2002.

\bibitem{Ste96}
A.~M. Steane.
\newblock Error-correcting codes in quantum theory.
\newblock {\em Phys. Rev. Lett.}, 77:793--797, 1996.

\end{thebibliography}
\bibliographystyle{plain}

\end{document}